\documentstyle[12pt,epsf]{article}
\newcommand{\ra}{\rangle}
\newcommand{\la}{\langle}

\newcommand{\II}{{\cal I}}

\newcommand{\AAA}{{\cal A}}
\newcommand{\MM}{{\cal M}}

\newcommand{\NN}{{\cal N}}

\newcommand{\wt}{\widetilde}

\newcommand{\bd}{{\bar {\rm D}}}

\newcommand{\be}{\begin{equation}}
\newcommand{\ee}{\end{equation}}
\newcommand{\ben}{\begin{eqnarray}\displaystyle}
\newcommand{\een}{\end{eqnarray}}
\newcommand{\refb}[1]{(\ref{#1})}
\newcommand{\p}{\partial}
\newcommand{\sectiono}[1]{\section{#1}\setcounter{equation}{0}}

\begin{document}

{}~ \hfill\vbox{\hbox{hep-th/9904207}\hbox{MRI-PHY/P990411}
}\break

\vskip 3.5cm

\centerline{\large \bf Non-BPS States and Branes in String Theory}  
\medskip

\vspace*{6.0ex}

\centerline{\large \rm Ashoke Sen
\footnote{E-mail: asen@thwgs.cern.ch, sen@mri.ernet.in}}

\vspace*{1.5ex}

\centerline{\large \it Mehta Research Institute of Mathematics}
 \centerline{\large \it and Mathematical Physics}

\centerline{\large \it  Chhatnag Road, Jhoosi,
Allahabad 211019, INDIA}

\vspace*{4.5ex}

\centerline {\bf Abstract}
We review the recent developments in our understanding 
of non-BPS states and branes in string theory. The topics include 1)
construction of
unstable non-BPS D-branes in type IIA and type IIB string theories, 2)
construction of stable non-BPS D-branes on various orbifolds and
orientifolds of type II string theories, 3) description of BPS and non-BPS
D-branes as tachyonic soliton solutions on brane-antibrane pair of higher
dimension,
and 4) study of the spectrum of non-BPS states and branes on a system of
coincident D-brane $-$ orientifold plane system. Some other related 
results are also discussed briefly.

\vfill \eject

\tableofcontents

\baselineskip=18pt

\sectiono{Introduction} \label{ss1}

In this article I shall review the recent progress in our 
understanding of stable non-BPS branes and states in string theory. 
These lectures will be based mainly on 
refs.\cite{9803194,9805019,9805170,9806155,9808141,9809111,9810188,9812031,9812135,9901014}.
We shall work in the convention $\hbar=1$, $c=1$, and $\alpha'=1$ 
(string tension=$(2\pi)^{-1}$) unless mentioned otherwise.

Let us begin with some motivation for studying non-BPS branes. There
are several reasons:
\begin{enumerate}

\item Stable non-BPS states and branes are very much part of the spectrum
of string theory, and our understanding of string theory remains
incomplete without a knowledge of these states.

\item Stable non-BPS states are the simplest objects whose masses
are not protected by supersymmetry, and yet are calculable in
different limits of the string coupling. Hence studying the spectrum
of these states in these different limits
might provide new insight into what happens at finite string
coupling.

\item A system of coincident non-BPS D-branes typically has, as
its world-volume theory, a non-supersymmetric gauge theory.
Thus they may be useful in getting results about non-supersymmetric
field theories from string theory, in the same way that a configuration
of supersymmetric branes can be used to study non-perturbative aspects
of supersymmetric gauge theories.

\item Non-BPS branes may be relevant for constructing string compactification 
with broken supersymmetry. 

\end{enumerate}

The plan of this article is as follows. In section \ref{ss2} we shall
discuss the construction of unstable non-BPS D-branes in type IIA and type
IIB string theories. Whereas type IIA (IIB) string theory admits stable
BPS branes of even (odd) dimensions, we shall see that they also admit
unstable non-BPS branes of odd (even) dimensions. In section \ref{ss3} we
shall show how on certain orientifolds / orbifolds of type II string
theories these non-BPS branes may give rise to stable non-BPS states and
branes. The main point here will be to note that under this orbifolding /
orientifolding operation the tachyonic mode responsible for the
instability of the non-BPS brane gets projected out. The resulting brane
is free from tachyonic mode and hence is stable. In section \ref{ss4} we
shall discuss the interpretation of the non-BPS branes in type IIA and IIB
string theories as tachyonic kink solution on a BPS D-brane - anti-D-brane
pair of one higher dimension in the same theory. We shall also show how
the BPS D-brane (anti-D-brane) can be regarded as a tachyonic kink
(anti-kink)
solution on a non-BPS D-brane of one higher dimension. This gives a set of
descent relations between BPS and non-BPS D-branes of type II string
theories, and form the basis of identifying the D-brane charge with
elements of
K-theory\cite{9810188,9812135,9812226,9901042,9902102,9902116,9902160,9904153}.
Since the actual proof of these relations is technically somewhat
complicated, we postpone the details to the appendix.

In section \ref{ss5} we discuss the spectrum of stable non-BPS states and
branes on a coincident D-brane and orientifold plane system. The masses
(tensions) of these states (branes) can be calculated in the strong
coupling limit using various duality symmetries of string theory. Although
for each system we use a completely different method for finding the
spectrum, the final spectrum of non-BPS states and branes on a D-$p$-brane
$-$ orientifold $p$-plane system exhibits an unusual regularity as a
function of $p$. Whether this signifies any deep aspect of 
string theory remains to be seen. Since the only similarity between
these systems is in their weak coupling perturbation expansion, we
suspect that the strong coupling result may be governed by large order
behaviour of this
perturbation expansion. Finally in section \ref{ss6}
we discuss some related developments in this subject. This includes a
discussion of some other non-BPS branes in type I string theory, 
the relationship between
K-theory and D-brane charges, and the application of boundary state
formalism in the study of non-BPS D-branes. We end with a discussion of
some open questions.

\sectiono{Unstable Non-BPS D-branes in Type II String Theories} 
\label{ss2}

\subsection{BPS D-branes in type II string theories}

\begin{figure}[|ht]
\begin{center}
\leavevmode
\epsfbox{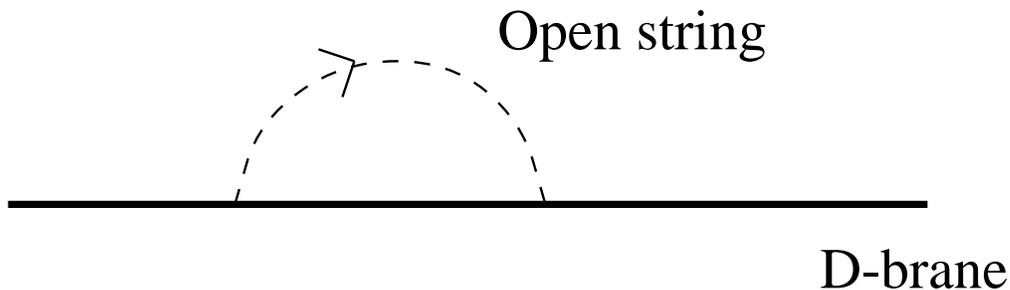}
\end{center}
\caption{Open strings ending on a BPS D-brane.} \label{f1}
\end{figure}

Let us begin by reviewing what we know about
BPS D-branes in type IIA/IIB string theories\cite{DBRANE}.
The defining property of the D-brane is that fundamental strings can
end on a D-brane as shown in Fig.\ref{f1}, although type II string
theories
in the bulk only 
contains closed string states without any end. The open strings with ends
on the D-brane can be interpreted as the dynamical modes of the D-brane.
In order to compute the spectrum of these open string
states with ends
on the D-brane, we impose Dirichlet boundary condition on the open string
coordinates along directions transverse to the D-brane,
and Neumann boundary condition along directions parallel to the
brane world-volume (including time).
A D-brane with $p$ tangential spatial
directions is called a D-$p$-brane.

Let us now list some of the
properties of D-branes in type II string theories which will be useful to
us later.
\begin{itemize}

\item Type IIA (IIB) string theory admits BPS D-$2p$-brane
(D-$(2p+1)$-brane) which are invariant under half of the
space-time supersymmetry transformations of the theory.

\item A D-$p$-brane is charged under a $(p+1)$-form gauge field
arising in the Ramond-Ramond (RR) sector of the theory.

\item These BPS D-branes are oriented.
D-branes of opposite orientation carry opposite RR charge
and will be called anti-D-branes ($\bar {\rm D}$-branes).

\end{itemize}
\begin{figure}[|ht]
\begin{center}
\leavevmode
\epsfbox{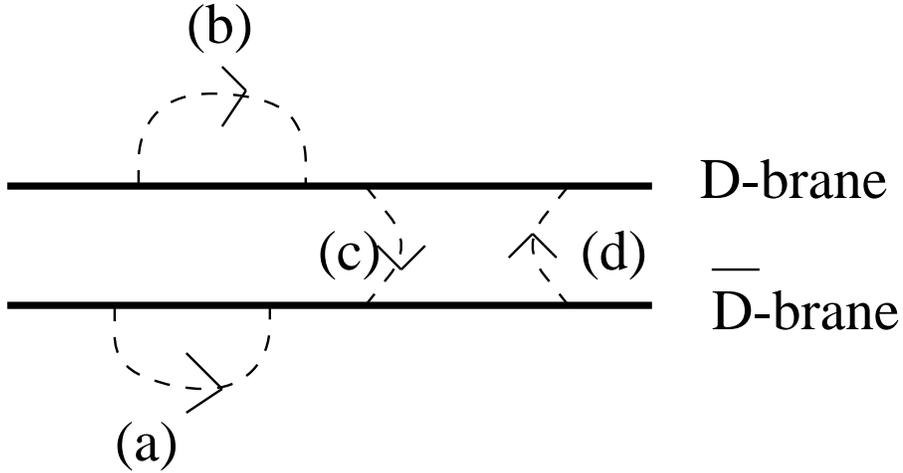}
\end{center}
\caption{Open strings living on a coincident D-brane anti-D-brane pair.
Although for clarity we have displayed the brane and the anti-brane to be 
spatially separated, we shall analyse the case where they coincide.}
\label{f2} \end{figure}

Next we shall review
properties of coincident D-brane $-$ $\bd$-brane pair shown in
Fig.\ref{f2}. They are as follows:
\begin{itemize}

\item
Spectrum of open strings living on the world-volume contains four different
sectors. These four sectors can be labelled by
$2\times 2$ Chan Paton (CP) factors:
\ben \label{es2.1}
&&
(a): \quad \pmatrix{0 & 0 \cr 0 & 1}, \qquad (b):\quad \pmatrix{ 1 & 0\cr
0 & 0} \nonumber \\
&&
(c):\quad \pmatrix{0 & 0 \cr 1 & 0}, \qquad (d):\quad \pmatrix{ 0 & 1\cr
0 & 0} \, .
\een

\item GSO projection: Physical states in sectors (a) and (b) should have
$(-1)^F=1$ whereas those in sectors (c) and (d) should have 
$(-1)^F=-1$.
Here $F$ denotes the {\it world-sheet fermion number} carried by the
state.  We use the convention that the $(-1)^F$ eigenvalue of the
Neveu-Schwarz (NS) sector
ground state is $-1$. The GSO projection
rule follows from the observation that the
closed string exchange interaction between a D-brane and a $\bd$-brane
and that between a pair of D-branes have the same sign for NSNS sector
closed string exchange and opposite sign for RR sector closed
string
exchange. In the open string channel this corresponds to replacing the
GSO projection operator ${1+(-1)^F\over 2}$ for DD strings by
${1-(-1)^F\over 2}$ for D$\bd$ strings.

\item Since the NS sector ground state has
$(-1)^F=-1$,
it survives the GSO projection 
in sectors (c) and (d) and gives tachyonic excitations
with\cite{9403040,9511194,9604091,9604156,9612215}
\be \label{es2.2}
m^2 = -(1/2)\, .
\ee
Since the tachyon comes from two different sectors it is a
complex scalar field.

\item Although individually the D-brane as well as the
$\bd$-brane is invariant under half of the space-time
supersymmetry transformations, the combined system breaks all
supersymmetries.

\end{itemize}

We shall now study the action of $(-1)^{F_L}$ on the coincident D-brane
$-$ $\bd$-brane system, where $F_L$ denotes the contribution to the {\it
space-time fermion number} from the left-moving sector of the string
world-sheet. $(-1)^{F_L}$ is known to be an exact symmetry of type IIA and
type IIB string theories. Acting on the closed string Hilbert space, it
changes the sign of all the states on the left-moving Ramond sector, but
does not change anything else. Thus it has trivial action on the
world-sheet fields. {}From this definition it follows that the space-time
fields originating in the Ramond-Ramond (RR) sector of the world-sheet
change sign under $(-1)^{F_L}$. Since D-branes are charged under RR field,
it follows that $(-1)^{F_L}$ must take a D-brane to a $\bd$-brane.
Thus a single D-brane or a single $\bd$-brane is not invariant under
$(-1)^{F_L}$, but 
a coincident D-brane $-$ $\bd$-brane system is invariant under
$(-1)^{F_L}$. Hence it makes sense to study the action of
$(-1)^{F_L}$ on the open strings living on this system, which is what we
shall do now.

We begin with the observation that since
$(-1)^{F_L}$ has no action on the world-sheet fields,
we only need to study its action on the CP factors.\footnote{We shall
focus our attention on the NS-sector states, but a similar analysis
can be done separately for the R-sector states.}
Since $(-1)^{F_L}$ exchanges D-brane with $\bd$-brane, it
acts on the CP matrix $\Lambda$ as
\be \label{es2.3}
\Lambda \to \sigma_1 \Lambda (\sigma_1)^{-1}\, ,
\ee
where
\be \label{es2.4}
\sigma_1 = \pmatrix{0 & 1\cr 1 & 0}\, .
\ee
This shows that states with CP factors $I$ and $\sigma_1$ are even under
$(-1)^{F_L}$, whereas those with CP factors $\sigma_3$ and $i\sigma_2$ are
odd. (We could
replace $\sigma_1$ by $\sigma_2$ in \refb{es2.3}, but this just amounts to 
a change in convention.)

\subsection{Non-BPS D-branes in type II string theories}

We are now ready to define a non-BPS D-$2p$-brane of type IIB
string theory\cite{9812031}.
This is done by following the steps listed below.
\begin{itemize}

\item
We start with a
D-$2p$ $-$ $\bd$-$2p$-brane pair in type IIA string theory and take the
orbifold of this configuration by $(-1)^{F_L}$.

\item
In the bulk, modding out IIA by $(-1)^{F_L}$ gives IIB.

\item
Acting on the open strings living on the D-$\bd$-brane
world-volume,
$(-1)^{F_L}$ projection keeps states with CP factors $I$ and
$\sigma_1$ and throws out states with CP factors $\sigma_3$ and
$i\sigma_2$.
\end{itemize}
This defines a {\it non-BPS D-$2p$-brane}
of type IIB string theory. In order to see that it describes a single
object rather than a pair of objects, we simply note that before the
projection the degree of freedom of separating the two branes reside in 
the sector with CP factor $\sigma_3$. Since states
in the CP sector $\sigma_3$ are projected out, we lose the degree of
freedom of separating the brane antibrane pair away from each other.
Similarly, starting from a D-$(2p+1)$-brane
$\bd$-$(2p+1)$-brane pair of IIB, and modding it out by $(-1)^{F_L}$, 
we can define a non-BPS
$(2p+1)$-brane of IIA.
Thus type IIB string theory contains BPS D-branes of odd dimension and
non-BPS D-branes of even dimension, whereas type IIA string theory
contains BPS D-branes of even dimension and non-BPS D-branes of odd
dimension.

Let us now list some of the
properties of the non-BPS D-$2p$-brane of type IIB string theory.
(Similar results also
hold for the non-BPS D-$(2p+1)$-brane of type IIA string theory.) These
properties
follow from their definition, and properties of coincident brane-antibrane
pair reviewed earlier.

\begin{itemize}
\item Excitations on its world-volume are open strings with
Dirichlet boundary condition on the $(9-2p)$ transverse
directions, and Neumann boundary condition on
$2p+1$ tangential directions (including time).

\item These open strings carry Chan Paton factors $I$ or
$\sigma_1$.

\item Physical states with CP factor $I$ has $(-1)^F=1$ and
physical states with CP factor $\sigma_1$ has $(-1)^F=-1$.
(Note again that $F$ denotes {\it world-sheet fermion number}.)

\item The NS sector ground state carrying CP factor $\sigma_1$ has
$(-1)^F=-1$ and hence is physical.
Thus there is
a tachyonic mode with 
\be \label{es2.5}
m^2 = -{1\over 2} \, .
\ee

\item Since tachyon comes from only one sector, it is a real scalar
field.

\item The tension of the non-BPS D-$2p$-brane of type IIB string theory 
is given by:
\be \label{es2.6}
(2\pi)^{-2p} (\sqrt 2/g)\, ,
\ee
where
$g$ denotes the coupling constant of the string theory. This property
can be derived by taking into account the effect of modding out by
$(-1)^{F_L}$, and the fact that the original brane-antibrane system before
$(-1)^{F_L}$ modding had a tension equal to
\be \label{es2.7}
(2\pi)^{-2p} (2/g)\, .
\ee
Similarly, the tension of a non-BPS D-$(2p+1)$-brane of type IIA string
theory is given by
\be \label{eiia}
(2\pi)^{-(2p+1)}(\sqrt 2/g)\, .
\ee

\end{itemize}

One can also derive the spectrum of open strings with one end on
the non-BPS brane and other end on a BPS brane, but we shall not
discuss it here.

\subsection{BPS D-branes from non-BPS D-branes}

Let us now
consider the effect of modding out a non-BPS D-$2p$-brane of
IIB by $(-1)^{F_L}$, where $(-1)^{F_L}$ now denotes the corresponding 
symmetry of the type
IIB string theory\cite{9812031}.
In the bulk, modding out type IIB string theory by $(-1)^{F_L}$ gives us 
back a type IIA string theory.
The question we shall be interested in is:
what happens to the D-$2p$-brane after this modding? This question
makes
sense as the non-BPS D-brane does not carry any RR charge and hence is
invariant under the action of $(-1)^{F_L}$.
In order to answer this question
we need to study the action of $(-1)^{F_L}$ on the
open string states living on the D-$2p$-brane.
As before $(-1)^{F_L}$ does not act on the world-sheet fields,
but acts only on the CP factors.
Thus we need to find the action of $(-1)^{F_L}$ on CP factors.
This is done with the help of the following
observations:\footnote{For definiteness we shall focus our
attention on the NS sector states, but a similar analysis can also be
carried out for R sector states.}
\begin{figure}[|ht] 
\begin{center}
\leavevmode
\epsfbox{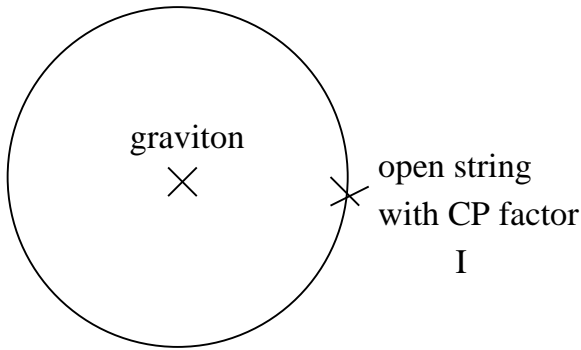}
\end{center}
\caption{The disk amplitude for two point function of the graviton and
translation mode of the D-brane.} \label{f3}
\end{figure}
\begin{itemize}
\item There is a non-zero two point function of the graviton $g_{m\mu}$ 
from the
closed string sector and the translation modes $X^m$ of the
D-$2p$-brane originating in the CP sector $I$ of the form:
\be \label{egfmet}
\eta^{\mu\nu} g_{m\mu} \p_\nu X^m \, ,
\ee
where $m$ denotes a direction transverse to the brane and $\mu,\nu$ denote
directions tangential to the brane. This coupling follows from expanding
the Dirac-Born-Infeld action on the brane world-volume around the
configuration of a flat brane in a flat space-time background.
This can also be seen by computing a disk amplitude with a graviton vertex
operator inserted at the center of the disk, and the tachyon vertex
operator inserted at the boundary of the disk, as shown in
Fig.\ref{f3}.\footnote{This does not mean that that a physical on-shell
scalar particle on the brane has a finite transition probability into a
graviton state in the bulk. This is disallowed due to various kinematic
reasons. However, the existence of the coupling \refb{egfmet} can still be
deduced by evaluation the disk amplitude in a region of unphysical
(complex) external momenta; as is done {\it e.g.} in  deducing the
Yang-Mill's three gauge boson vertex from three string
amplitude\cite{SCHPHYSREP}.}
Since the graviton is even under $(-1)^{F_L}$, this shows that states with
CP
factor $I$ must also be even under $(-1)^{F_L}$.
\begin{figure}[|ht]
\begin{center}
\leavevmode
\epsfbox{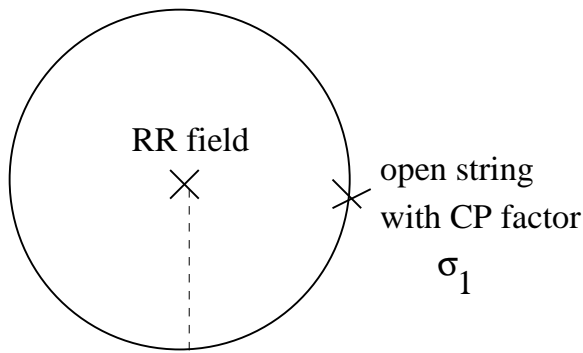}
\end{center}
\caption{The disk amplitude for the two point function of the tachyon and
the RR sector $2p$-form gauge field. The dotted line denotes the
$(-1)^{F_L}$ cut
extending from the RR vertex operator to the disk boundary.}
\label{f3a}
\end{figure}

\item There is a non-zero two point function of the RR-sector
$2p$-form gauge field $A^{(2p)}$ from the closed string sector and the
tachyonic
mode $T$
of the D-$2p$-brane originating in the CP sector $\sigma_1$ of the form:
\be \label{etachrr}
\int A^{(2p)}\wedge dT\, .
\ee
This can be seen by computing the disk amplitude with a RR-sector gauge
field vertex operator inserted at the center of the disk, and the tachyon
vertex operator inserted at the boundary, as shown in
Fig.\ref{f3a}.\footnote{Again, as before, the
actual transition between a massless RR sector state
and the tachyon is
absent due to kinematic reasons.} The fact that this amplitude
is non-zero may seem surprising, as the tachyon vertex
operator carries a CP factor $\sigma_1$, and there seems to be no other CP
factor inserted at the boundary of the disk. However, since the RR sector
states in type IIB string theory appear in the twisted sector when we
regard type IIB string theory as type IIA string theory modded out by
$(-1)^{F_L}$, there is a cut extending from the RR sector vertex operator
at the center all the way to the boundary of the disk. At the point where
the cut hits the boundary we need to insert an extra factor of $\sigma_1$,
since $(-1)^{F_L}$ action on the CP factors correspond to conjugation by
$\sigma_1$. This gives a total of two factors of $\sigma_1$ on the
disk boundary and makes the amplitude non-vanishing. {}From this it
follows
that since RR-sector fields are odd under $(-1)^{F_L}$, states with
CP factor $\sigma_1$ are also odd under $(-1)^{F_L}$.

\end{itemize}

The
net result of this analysis is that states with CP factor $I$ are
$(-1)^{F_L}$ even and states with CP
factor $\sigma_1$ are $(-1)^{F_L}$ odd.
Thus under modding out by $(-1)^{F_L}$, only states with CP
factor $I$ survive the projection.
As we have already seen before,
GSO projection requires these states to be even under $(-1)^F$.
Thus the spectrum is identical to that of open strings living on
a {\it BPS D-$2p$-brane of IIA}, and we conclude that
the non-BPS D-$2p$-brane of type IIB string theory,
modded out by $(-1)^{F_L}$, gives a BPS D-$2p$-brane of
type IIA string theory.\footnote{Note that at this
stage we cannot determine whether the resulting brane is a D-brane or a
$\bd$-brane, as both carry the same spectrum of open string. This is a
reflection of the fact that the orbifolding procedure has a two-fold
ambiguity, so that we could end up either with a D-brane or a $\bd$-brane
by following these steps.}
\begin{figure}[|ht]
\begin{center}
\leavevmode
\epsfbox{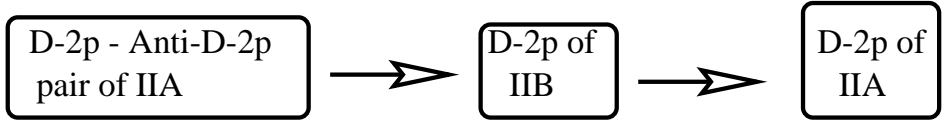}
\end{center}
\caption{Relationship between BPS and non-BPS D-branes in type II
string
theories. 
The horizontal arrow
represents the effect of modding out the theory by
$(-1)^{F_L}$.} \label{f4}
\end{figure}

The results of this section have been summarized in Fig.\ref{f4}.
There is also a similar relation with
IIA $\leftrightarrow$ IIB
and 
$(2p)\rightarrow(2p+1)$.

\sectiono{Stable Non-BPS D-branes on Type II Orbifolds and Orientifolds} 
\label{ss3}

Although we have constructed non-BPS D-branes in type IIA/IIB string
theory in the last section, they are all unstable due to the presence of
the tachyonic mode.
As we shall discuss in section \ref{ss4}, 
if the tachyon condenses to its minimum, then the configuration
is indistinguishible from the vacuum\cite{9805170}. 
Thus it is natural to ask:
what is the use of such a D-brane?

In this section we shall show that
although they are unstable in type
IIA/IIB string theory, we may get
{\it stable} non-BPS D-branes in certain orbifolds/orientifolds of
IIA/IIB {\it if the tachyonic mode is projected out under this
operation.}
We shall illustrate this through two examples.

\subsection{Type I D-particle} \label{stypei}

Let us consider the following construction:
\begin{itemize}
\item Start with the non-BPS D0-brane
(D-particle) of type IIB as defined in the last section.

\item Mod out the configuration by the world-sheet parity
transformation $\Omega$.

\end{itemize}

The result can be described as a
non-BPS D-particle of type I string theory, since in the bulk type IIB
string theory modded out by $\Omega$ gives a type I string theory.
The crucial question is:
is this D-particle stable?
Or equivalently we may ask:
is the tachyonic mode on the type IIB D-particle odd under
$\Omega$?
The answer to this question follows from eq.\refb{etachrr} for $p=0$,
{\it i.e.} 
that the two point function of the tachyonic mode on the
D-particle world-volume and the RR sector scalar field $\phi$ of
type IIB string theory is non-vanishing. 
Since the field
$\phi$ is known to be odd under $\Omega$, we conclude that
the tachyonic mode of the D-particle is also odd under
$\Omega$.
Thus it is projected out in type I string theory.
In other words, the type I D-particle is stable\cite{9809111}!

The spectrum of open strings on type I D-particle also includes open
strings with one end on the D-particle and the other end on any one of the
32 nine branes which are present in type I string theory. The Ramond
sector states from this sector can be shown to give rise to 32 massless
fermionic zero modes living on the D0-brane. Quantization of these zero
modes gives rise to a ground state which transforms in the spinor
representation of the type I gauge group SO(32). This also gives an
additional explanation of the stability of the D-particle. Since all
perturbative states of type I string theory are in the scalar conjugacy
class of SO(32), and since a spinor state cannot decay into states in the
scalar conjugacy class, the D-particle is prevented from decaying into
perturbative string states due to charge conservation.

If we consider two or more coincident D-particles in type I string theory,
then there are also possible tachyonic modes coming from open strings with
two ends on two different D-particles. It turns out that the $\Omega$
projection does not remove all the tachyonic modes from these sectors, and
two or more coincident D-particles describe an unstable system. This is
consistent with the observation that two particles in the spinor
representation of SO(32) can combine and annihilate into
perturbative string states, as there is no conservation law preventing
this process.

The existence of the type I D-particle is also relevant for testing the
conjectured
duality between type I and heterotic string
theory\cite{9503124,9506160,9506194,9510169}.
SO(32) heterotic string theory contains states in the
perturbative spectrum which transform in the spinor
representation of SO(32).
These states are massive, and non-BPS.
But the lightest state belonging to the spinor representation of
SO(32) is stable at all values of the coupling,
as they cannot decay into anything else. Thus these states must also exist
in the strong coupling limit of the SO(32) heterotic string theory, which
is nothing but the weakly coupled type I string theory. The type I
D-particles provide explicit realization of these states.

It is instructive to compare the mass formulae for these SO(32)
spinor states at the two extreme ranges of the coupling constant. We shall
use the variables of the heterotic string theory to express this mass
formula at the two ends. 
For small heterotic coupling $g_H$, the 
perturbative mass formula in the heterotic string theory holds:
\be \label{es3.1}
\sqrt{T_H}\, (a_0+a_1\, g_H^2+a_2\, g_H^4+\ldots)\, ,
\ee
where $T_H$ is the heterotic string tension, and $a_i$ are numerical
coefficients. $a_0$ is computed at tree level of heterotic string theory,
whereas $a_m$ is computed at $m$-loop order.

For large heterotic coupling we can use the description of this state as
type I D-particle to compute its mass.
As we saw earlier, this has mass of order
$\sqrt{T_I}/g_I$, where $T_I$ and $g_I$ are the string tension and
coupling constant respectively of the type I string theory. 
Using standard relationship between the heterotic and type I
variables\cite{9503124}
\be \label{es3.2}
T_I = T_H g_H^{-1}, \qquad g_I=g_H^{-1}\, ,
\ee
we see that for large $g_H$ the mass of this state is proportional to:
\be \label{es3.3}
\sqrt{T_H}  (g_H)^{1/2}\, .
\ee
It will be interesting to see if the perturbation expansion \refb{es3.1}
contains any information about the large $g_H$ behaviour given in
\refb{es3.3}.

\subsection{D-branes wrapped on non-supersymmetric cycles of K3 orbifold}

In this section we shall discuss another example where the tachyonic mode
of a non-BPS D-brane is projected out under an orbifolding
operation\cite{9812031}. We
proceed as follows:
\begin{itemize}
\item
Start with a non-BPS D-string of type IIA string theory wrapped on a
circle along $x^9$
of radius $R_9$ and placed at $x^i=0$ for $1\le i\le 8$.
\item
Compactify three other directions $x^6,x^7,x^8$.
\item 
Mod out the theory by a $Z_2$ transformation $\II_4$ which
changes the sign of $x^6,\ldots x^9$:
\be \label{es3.4}
\II_4: \quad (x^6,x^7,x^8,x^9)\to (-x^6,-x^7,-x^8,-x^9)\, .
\ee
\end{itemize}

In the bulk this gives type
IIA string theory on an orbifold K3. We shall now analyze the fate of the
tachyon field
on the D-string in this orbifold theory. 
Since the D-string lies along $x^9$, the tachyon field on its
world-steet is a function of $x^9$ and time $t$. Again by considering a
two point function between the tachyon and 
an RR sector gauge field, one
can show that under the $Z_2$ transformation $\II_4$,
\be \label{es3.5}
T(x^9,t)\to - T(-x^9,t)\, .
\ee
If we expand $T(x^9,t)$ in its Fourier mode as:
\be \label{es3.6}
T(x^9,t) = \sum_n T_n(t) e^{inx^9/R_9}\, ,
\ee
then under $\II_4$:
\be \label{es3.7}
T_n(t) \to - T_{-n}(t)\, .
\ee
Thus
\begin{itemize}
\item
$T_0$ is projected out.
\item For $n\ne 0$ the combination $T_n - T_{-n}$ survives the
projection under $\II_4$.
\end{itemize}
Since the 
effective mass$^2$ of $T_n-T_{-n}$ is given by
\be \label{es3.8}
m_n^2 = (n^2/R_9^2)-(1/2)\, ,
\ee
we see that
there is no tachyon in the spectrum for
\be \label{es3.9}
R_9\le \sqrt 2\, .
\ee

There are also possible tachyonic modes from open string states stretched
between the original D-string and its image under translation along $x^6$,
$x^7$ or $x^8$.
Demanding that there are no tachyonic modes from these sectors
we also get\footnote{These relations can be found from 
eq.\refb{es3.9} by a
T-duality transformation.}
\be \label{es3.10}
R_8\ge{1\over \sqrt 2}, \quad R_7\ge {1\over \sqrt 2},
\quad R_6\ge {1\over \sqrt 2}\, .
\ee
\begin{figure}[|ht]
\begin{center}
\leavevmode
\epsfbox{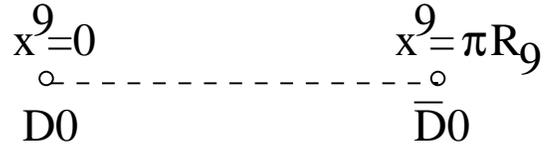}
\end{center}
\caption{The D0-$\bd$0 brane configuration obtained by marginal
deformation of the non-BPS D-string wrapped along $x^9$.} \label{f8}
\end{figure}

The net result of this analysis is that we have
a stable non-BPS state in type IIA string theory on $T^4/\II_4$ in the
range of
parameters described in \refb{es3.9}, \refb{es3.10}. The next question
would be:
what is the interpretation of this state?
There are many ways to answer the question; we shall explain it by 
studying the physics at the critical radius
$R_9=\sqrt 2$.
At this radius $T_{\pm 1}$ are massless modes.
In fact one can show that the potential for $(T_1-T_{-1})$
vanishes identically.\footnote{This and various other issues related to
this discussion will be discussed in some detail in section \ref{ss4}.} 
Thus $(T_1-T_{-1})$ denotes an exactly marginal deformation of
the boundary conformal field theory (CFT) describing the D-brane.
We can study this deformation using CFT techniques.
We shall only quote the result here (see section \ref{ss4}, the appendix 
and
ref.\cite{9812031} for some of the details). It turns out that
this marginal deformation takes the non-BPS D-string of
IIA to a D$0$-$\bd$0-brane pair situated at the two
fixed points $x^9=0$ and $x^9=\pi R_9$ respectively as shown in
Fig.\ref{f8}.

So far what we have described could have been done even before modding
out the
theory by $\II_4$. Let us now study the result of modding out this
configuration by $\II_4$.
It was shown in ref.\cite{9603167}
that after modding out by $\II_4$ a
$D0$-brane at $x^9=0$ can be interpreted as
a D2-brane of type IIA string theory,
wrapped on the supersymmetric
2-cycle\cite{9507158} associated with the fixed point of $\II_4$ at
$x^9=0$.\footnote{Although the cycle has zero area, the wrapped D-brane
has a finite mass due to the presence of the anti-symmetric tensor field
flux through the two cycle\cite{9507012}.} A similar interpretation can be
given for the $\bd$0-brane
at $x^9=\pi R_9$. Thus in the orbifold theory the marginal perturbation by
$(T_1-T_{-1})$ at $R_9={\sqrt 2}$ takes the original non-BPS state
to a pair of D2-branes, wrapped on the 2-cycles associated with the
fixed points at $x^9=0$ and $x^9=\pi R_9$ respectively. This suggests
that the original configuration
is a D-2-brane of IIA wrapped simultaneously on both these 2-cycles.
This represents a D2-brane wrapped on
a non-supersymmetric 2-cycle.

Before the projection, the mass of the wrapped non-BPS D-string is given
by $(\sqrt 2 R_9/g)$, whereas the sum of the masses of the D0-$\bd$0 pair
is given by $(2/g)$. Modding out by $\II_4$ reduces the mass of each
state to half its original value.
By comparing the masses of the various (wrapped) branes
we arrive at the following picture:
\begin{itemize}
\item At the critical radius the D-2-brane
wrapped on the
non-supersymmetric cycle is degenerate with the pair of
D-2-branes wrapped on the supersymmetric cycles.

\item Below the critical radius the D-2-brane
wrapped on the
non-supersymmetric cycle is lighter than the pair of
D-2-branes wrapped on the two supersymmetric cycles.
Hence this wrapped brane is stable.

\item Above the critical radius the D-2-brane
wrapped on the
non-supersymmetric cycle is heavier than the pair of
D-2-branes wrapped on the two supersymmetric cycles.
As a result
this wrapped brane is unstable against decay into a pair of
supersymmetric brane configurations.

\end{itemize}

This construction can be generalized to describe a $(2p+2)$-brane
($(2p+1)$-brane) of IIA (IIB)
wrapped on a non-supersymmetric cycle of K3.
Using this procedure one can also construct examples of D-branes
wrapped on non-BPS 2- and 3-cycles of Calabi-Yau manifolds. These
generalizations have been discussed in ref.\cite{9812031}.

Before concluding this discussion we note that the
world-volume theory of $N$ coincident branes of this type gives
rise to a non-supersymmetric U(N) gauge theory.
This might be useful in solving non-supersymmetric field
theories via branes.

\sectiono{D-branes as Tachyonic Kink Solutions} \label{ss4}

\subsection{Non-BPS D-brane as tachyonic kink on the brane-antibrane
pair}
\begin{figure}[|ht]
\begin{center}
\leavevmode
\epsfbox{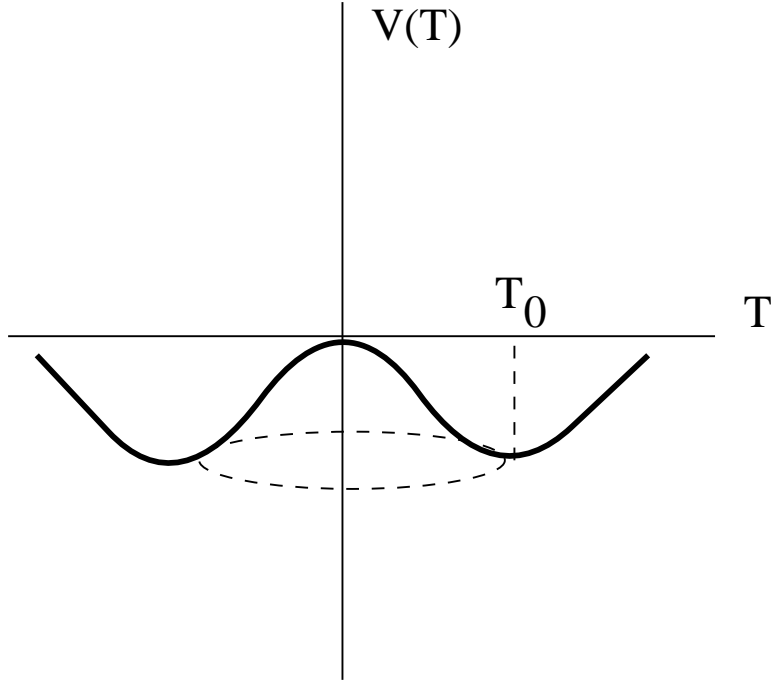}
\end{center}
\caption{The tachyon potential on D-brane $-$ $\bd$-brane pair.}
\label{f5}
\end{figure}

In this section we shall  give an alternative construction of the
non-BPS D-branes discussed in section \ref{ss2}\cite{9805019,9808141}.
Let us start with a
coincident pair of D-$2p$ $-$ $\bd$-$2p$ branes ($p\ge 1$) of type IIA
string theory.
As discussed in section \ref{ss2}, there is a complex tachyon field $T$
living on the world-volume of this
system.
This reflects the fact that $T=0$ is the maximum of the tachyon
potential $V(T)$ obtained after integrating out all other massive modes
on the world-volume. There is a U(1)$\times$U(1) gauge field living on the
world-volume of the brane-antibrane system, and the tachyon picks up
a phase under each of these U(1) gauge transformations. As a result,
$V(T)$ is a function only of $|T|$, and
the minimum of the potential occurs at $T=T_0 e^{i\theta}$ for some fixed
$T_0$ but arbitrary $\theta$, as shown in Fig.\ref{f5}.
As we shall argue shortly,
at the minimum, the sum of the tension of the
D-brane $\bd$-brane pair and the (negative) potential energy of
the tachyon is {\it exactly zero\cite{9805170} i.e.}, 
\be \label{es4.1}
2 T_D + V(T_0) = 0\, ,
\ee
where
$T_D$ is the D-brane tension.
This shows that the tachyonic ground state $T=T_0$
is indistinguishible from the vacuum, since it carries neither any charge 
nor any energy density.

\begin{figure}[|ht]
\begin{center}
\leavevmode
\epsfbox{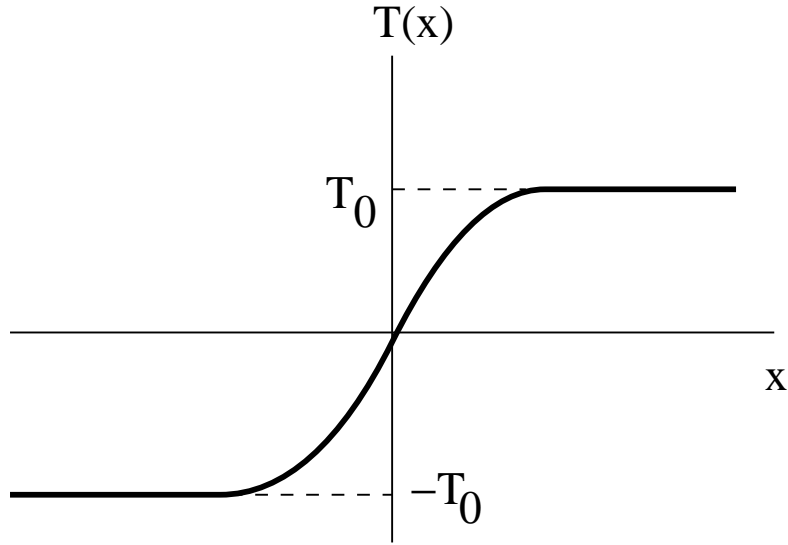}
\end{center}
\caption{Tachyonic kink solution on the brane-antibrane pair.} \label{f6}
\end{figure}
But now, instead of considering tachyonic
ground state, let us
consider a tachyonic kink solution.
For this,
consider the minimum energy configuration with the following properties:
\begin{itemize}
\item $Im(T)=0$.
\item $Re(T)$ independent of time and $(2p-1)$ of the $2p$ spatial
coordinates.
\item $Re(T)$ depends on the remaining spatial coordinate $x$
such that
\ben \label{es4.2}
T(x) \to T_0 \quad &\hbox{as}& \quad x\to \infty, \nonumber \\
T(x)\to -T_0 \quad &\hbox{as}& \quad x\to -\infty\, .
\een
\end{itemize}
This has been shown in Fig.\ref{f6}.
{}From this it is clear that
as $x\to \pm\infty$ the solution goes to vacuum configuration.
Thus the energy density is concentrated around a $(2p-1)$ dimensional
subspace, and the solution describes a $(2p-1)$-dimensional brane. We now
claim that
{\it
this $(2p-1)$-brane associated with the tachyonic kink solution on the
brane-antibrane pair is identical to the non-BPS
D-$(2p-1)$-brane of IIA.}

Note that $V(T)$ cannot be explicitly calculated.
Thus one might ask
how one could show the equivalence between the non-BPS D-brane described
in section \ref{ss2} and
the tachyonic kink on the brane $-$ antibrane pair described here. This
will be discussed in some detail in the appendix; but here we shall
describe the
outline of the proof.

\begin{itemize}

\item
There is a marginal deformation involving bulk and boundary
operators which interpolates between the $T=0$ configuration and
the kink solution.

\item One can study the fate of the CFT describing the
brane-antibrane pair under this marginal deformation.

\item The end result turns out to be a CFT which is identical to
the CFT describing the non-BPS D-brane.

\end{itemize}

One can also give an intuitive understanding of why a tachyonic kink
should behave like a D-brane. For this note that far away from the kink
(large $|x|$) the configuration represents the vacuum, and hence strings
cannot end there. On the other hand, on the subspace $x=0$, the tachyon
field vanishes, and hence we expect the configuration to behave in a way
that a D-brane $-$ $\bd$-brane pair would have behaved in the absence of
tachyon vev, {\it i.e.} open strings should be able to end there. Thus the
tachyonic kink should at least qualitatively behave as a D-brane located
at $x=0$.

Note that the manifold $\MM$ describing the minimum of the tachyon
potential is a circle $S^1$.
In order to get a topologically stable kink solution, we need 
$\pi_0(\MM) \ne 0$.
But $\pi_0(S^1)=0$ since $S^1$ is connected.
Thus the kink is not topologically stable.
Indeed it has tachyonic mode correponding to the freedom of
changing $T$ at $x\to\infty$ to $T_0 e^{i\theta}$.
As $\theta\to\pi$ we get back the vacuum configuration, since $T\to
-T_0$ as $x\to\pm\infty$ in this case.
This however is completely consistent with the identification of this kink
solution with the non-BPS D-$(2p-1)$-brane of type IIA string theory,
since, as we have seen earlier, the latter also has a tachyonic mode
living on it. 
The tachyonic mode on the kink solution can be identified as the
tachyonic mode on
the non-BPS D-$(2p-1)$-brane of IIA discovered in section \ref{ss2}.

Before we move on to the next subject, let us give an argument in favour
of eq.\refb{es4.1}. For this, note that if \refb{es4.1} had not been true,
then the tachyonic kink solution described here will not have a finite
energy per unit $(2p-1)$-volume, since the energy density, integrated
along
the transverse direction (denoted by $x$ in eq.\refb{es4.2}) would give
infinite answer. On the other hand from the analysis of section \ref{ss2} 
we certainly know that a non-BPS
D-$(2p-1)$ brane of type IIA string theory has finite tension. Thus once
we establish the equivalence of the tachyonic kink solution and the
non-BPS D-brane (as will be discussed in the appendix), it
automatically establishes eq.\refb{es4.1}. 

\subsection{The BPS D-brane as the tachyonic kink on the non-BPS D-brane}
\begin{figure}[|ht]
\begin{center}
\leavevmode
\epsfbox{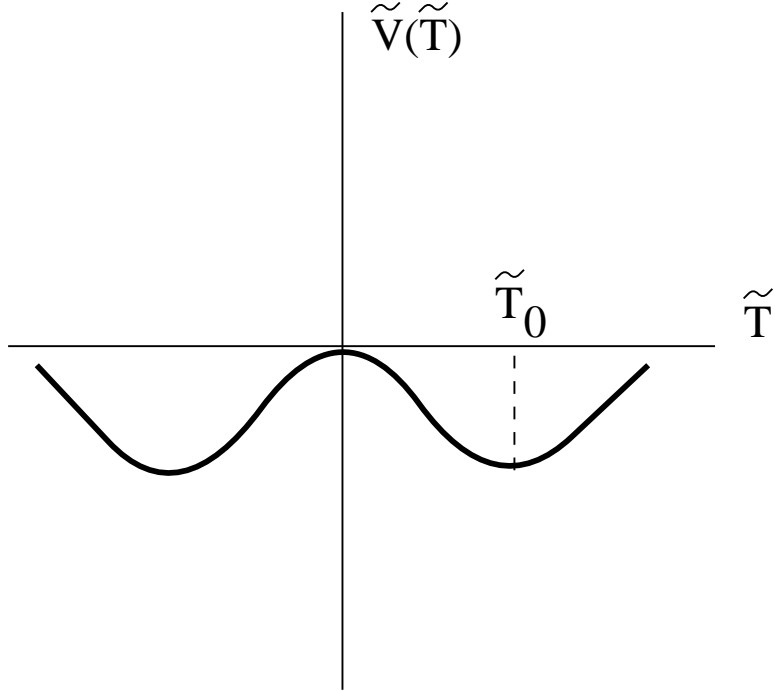}
\end{center}
\caption{The tachyon potential on the non-BPS D-brane} \label{f5a}
\end{figure}

We can now continue one step further.
Let us
start with a non-BPS D-$(2p-1)$-brane of IIA.
As was discussed in section \ref{ss2}, it has a real tachyon $\wt T$.
By studying the disk amplitude it can be easily seen that there is a $Z_2$
symmetry on the world-volume of this non-BPS D-brane under which $\wt
T$ (and all other modes originating in the CP
sector $\sigma_1$) changes sign. Let $\pm \wt T_0$ be the minimum of the
tachyon potential $\wt V(\wt T)$ obtained after integration out the other
massive modes, as shown in Fig.\ref{f5a}.
Again one can argue that: 
\be \label{es42.1}
\wt V(\wt T_0) + \wt T_D =0\, ,
\ee
where $\wt T_D$ is the tension of the non-BPS D-brane. 
We now
consider a kink solution on this D-$(2p-1)$-brane world-volume
such that:
\begin{figure}[|ht]
\begin{center}
\leavevmode
\epsfbox{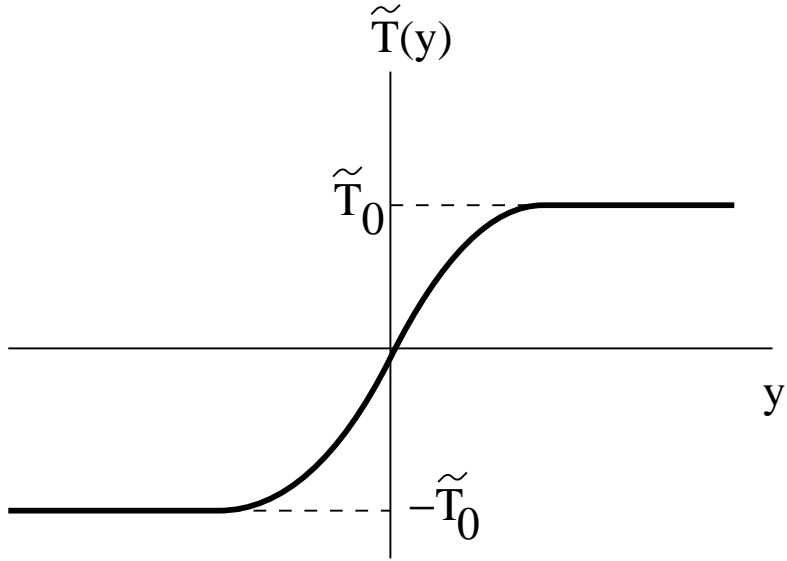}
\end{center}
\caption{Tachyonic kink solution on a non-BPS D-brane.} \label{f6a}
\end{figure}

\begin{itemize}
\item
$\wt T$ is independent of time as well as $(2p-2)$ of the spatial
coordinates.

\item It depends on the remaining world-volume coordinate $y$
such that:
\ben \label{es42.3}
\wt T(y)\to \wt T_0 \quad &\hbox{as}& \quad y\to \infty, \nonumber \\
\wt T(y)\to -\wt T_0 \quad &\hbox{as}& \quad y\to -\infty\, .
\een

\end{itemize}
This configuration has been shown in Fig.\ref{f6a}.
By the same argument as in the previous subsection, this describes a
$(2p-2)$ dimensional brane. We shall show in the appendix that this
can be identified as the BPS D-$(2p-2)$ brane of type IIA string theory.
The analysis is again based on finding a series of marginal deformations
involving bulk and boundary operators which connect
the $\wt T=0$ configuration of the non-BPS D-$(2p-1)$ brane to a
solution representing a kink-antikink pair,
and using conformal field theory techniques to 
show that this marginal deformation actually interpolates
between the non-BPS D-$(2p-1)$-brane and a BPS
D-$(2p-2)$-brane $-$ $\bd$-$(2p-2)$-brane pair.

Note that now the manifold
$\wt\MM$ describing the minimum
of the tachyon potential consists of a pair of points $\pm \wt T_0$.
Thus $\pi_0(\wt\MM) \ne 0$, and hence
the kink is stable as is expected of a BPS D-brane. An argument similar to
the one in the previous subsection can be used to give an intuitive
explanation of why the kink should behave as a D-brane near $y=0$ but
as vaccuum for large $|y|$. We can also explain the origin of the RR
charge
of the kink from the coupling \refb{etachrr} and the
fact that $\p_y T$ is non-zero at $y=0$. Since $\p_y T$ has opposite sign
for the anti-kink, this also shows that  the anti-kink must represent the
BPS $\bd$-$(2p-2)$ brane.
\begin{figure}[ht]
\begin{center}
\leavevmode
\epsfbox{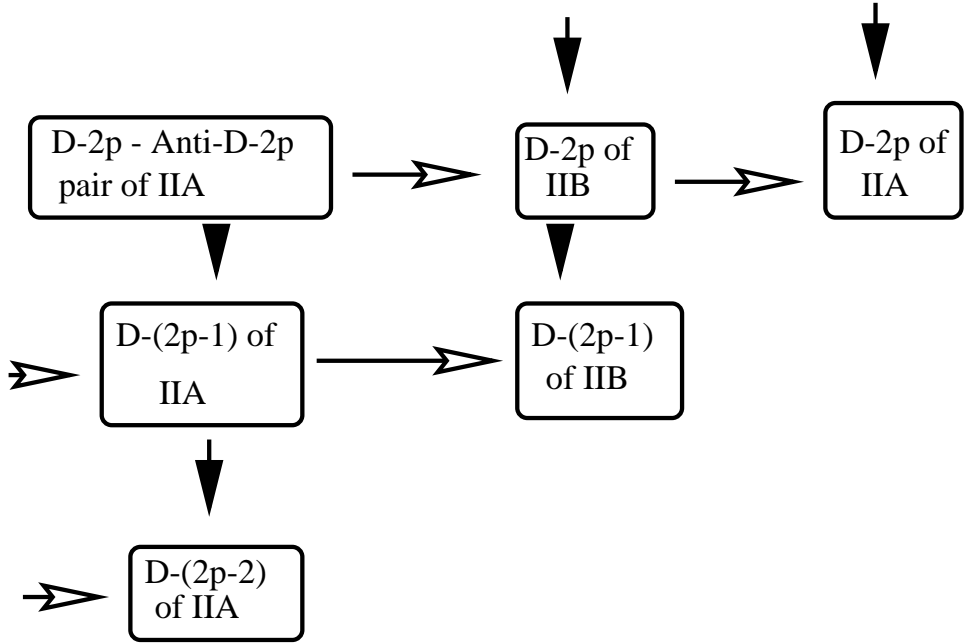}
\end{center}
\caption{Descent relations among BPS and non-BPS D-branes in type II
string theories.
The horizontal arrows denote the effect of modding out by $(-1)^{F_L}$,
and the
vertical arrows denote the ffect of considering tachyonic kink solution.}
\label{f7}
\end{figure}

The results of this section, combined with the results of section
\ref{ss2} leads to the
set of `descent relations' between BPS and
non-BPS D-branes shown in Fig.\ref{f7}.

By combining the two main results of this section, we can also
represent a BPS D-$p$-brane as a soliton (vortex) solution on the
D-$(p+2)$-brane $-$ $\bd$-$(p+2)$-brane pair in the same
theory\cite{9808141,9810188}. This construction is relevant for
relating allowed D-brane charges to elements of the K-group of
space-time\cite{9810188}. 

\sectiono{Stable Non-BPS Branes on the D-brane $-$ 
Orientifold Plane System} \label{ss5}

\subsection{Summary of the results} \label{s1}

We have already introduced the notion of a D-$p$-brane in type II string
theory. We now introduce the concept of an orientifold $p$-plane
(O-$p$-plane)\cite{ORIENT,9601038}. For this we consider
type II string theory on $R^{p+1}\times
(R^{9-p}/\II_{9-p}\cdot\Omega\cdot g)$, where
$\II_{9-p}$ reverses the sign of all the coordinates on $R^{9-p}$, 
$\Omega$ is the world-sheet parity transformation (L $\leftrightarrow$
R) and 
$g$ is  identity for $(9-p)=4m$ or $(4m+1)$
and
$g=(-1)^{F_L}$ for $(9-p)=(4m+2)$ or $(4m+3)$.
One can show that
$\II_{9-p}\cdot\Omega\cdot g$ is a symmetry transformation of order 2 in
type IIA string theory 
if
$p$ is even, and in type IIB string theory if
$p$ is odd.
The
origin of $R^{9-p}$ will be called an orientifold $p$-plane.
Thus type IIA (IIB) string theory contains orientifold $p$-planes of
even (odd) dimensions.

\begin{figure}[|ht]
\begin{center}
\leavevmode
\epsfbox{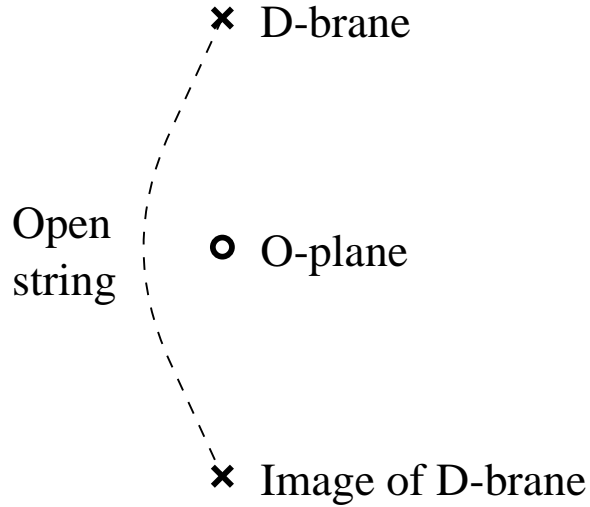}
\end{center}
\caption{Transverse section of the
coincident D-$p$-brane $-$ O-$p$-plane system. Although for
clarity we have
shown the D-brane and the O-plane as separated in space, we shall analyze 
the case where they are on top of each other.} \label{f13}
\end{figure}
Our focus of attention in this section will be a system of parallel
D-$p$-brane $-$ O-$p$-plane system. This corresponds to starting with a
D-brane and its
image under $\II_{9-p}\cdot\Omega\cdot g$, and then modding out the
theory by $\II_{9-p}\cdot\Omega\cdot g$, as shown in Fig.\ref{f13}.
The world volume theory of a
Dirichlet $p$-brane (D-$p$-brane) on top
of an orientifold $p$-plane (O-$p$-plane) has as its low energy limit an
$N=4$ supersymmetric SO(2) gauge theory.\footnote{There is some
ambiguity in how we choose the action of this $Z_2$
transformation on the CP factors; and due to this ambiguity we
can get different kinds of orientifold 
planes\cite{9601038,9712028}. 
Throughout
this paper we
shall only consider orientifold planes of SO-type $-$ also known as
the O$^+$ planes\cite{9712028}
$-$ carrying negative RR charge compared to that
of a D-brane.} The spectrum of stable
states in this theory contains a massive non-BPS state carrying unit
charge under this SO(2) gauge field.  These arise from open strings
stretched between the D-brane and its image.\footnote{Before the
orientifold
projection the ground state in this sector is massless and corresponds to
the charged vector bosons and their superpartners, but the orientifold
projection removes this state from the spectrum.} In the weak coupling
limit
these states have mass of the order of the string scale $m_S$ with
corrections expressible as a perturbation series in the string
coupling $g_S$: 
\be \label{eperturb}
m_S(K_0+K_1 g_S + K_2 g_S^2 +\cdots)\, .
\ee
Here $K_0,K_1,K_2,\ldots$ are numerical constants, with $K_m$ computed
from a diagram with $m$ open string loops.
Since the lowest
mass state carrying SO(2) electric charge must be stable at all values of
the
string coupling, it makes sense to ask what would be the masses of these
states in the strong coupling limit. This is one of the questions we
address in this section.
The answers were obtained in
refs.\cite{9803194,9805019,9806155,9808141}
and have been summarised in table 1.
\begin{center} 
\begin{tabular}{|c|c|c|} 
\hline $p$ & mass & $C_p$
\\ 
\hline 6 & $C_6 m_S g_S$ & known \\ 
\hline 5 & $C_5 m_S (g_S)^{1\over 2}$
& known \\ 
\hline 4 & $C_4 m_S (g_S)^{1\over 3}$ & unknown \\ 
\hline 3 &
unknown & $-$ \\ 
\hline 
\end{tabular} 
\medskip

Table 1: Masses of electrically charged states on the D-$p$-brane 
O-$p$-plane system in the strong coupling limit.
\end{center}

In this table, the first column denotes the value of $p$, the second
column denotes the mass of the lightest stable electrically charged state
on the D-$p$-brane $-$ O-$p$-plane system, $C_p$
denotes a numerical constant, and the last column denotes whether the
numerical constant $C_p$ is known or unknown at present. We have
restricted $p$ in the range $3\le p\le 6$ due to the following reason.
For $p\ge 7$, the dilaton does not go to a constant value
asymptotically\cite{9605150}, and as a result the string coupling $g_S$ is
not a
well defined quantity. On the other hand, for $p\le 2$, the self-energy of
an electrically charged particle blows up due to the long range Coulomb
field associated with the particle, and hence the mass of such a state is
not a well defined quantity.   

We shall review the arguments leading to these results in subsection
\ref{s2}. As we can see from this table, we still do not know the mass of
the
electrically charged particle on the D-3-brane $-$ O-3-plane system in the
strong coupling limit. Although it may be somewhat premature to look for a
pattern among three data points, we note that there seems to be some
regularity in the dependence of this mass on $g_S$ for $4\le p\le 6$,
namely it seems to go as
\be \label{e1}
m_S (g_S)^{1\over 7-p}\, .
\ee
Considering that for different values of $p$ these results are derived
using very different techniques, one might wonder if there is a deeper
lesson about strongly coupled string theory in this spectrum. Since the
only feature that is common between different values of $p$ is the
structure of weak coupling perturbation theory, it is tempting to
speculate that the regularity of the strong coupling spectrum is a
reflection of the regularity of the weak coupling perturbation theory
as a function of $p$. In that case we can expect that the information
about the strong couping
result is somehow contained  in the weak coupling perturbation theory,
$-$
in particular in its large order behaviour.

Besides stable non-BPS states which are electrically charged under the
SO(2), the brane world-volume theory also contains branes which are
magnetically charged under the SO(2). On the D-$p$-brane O-$p$-plane
system these are $(p-3)$ branes, and come from a D-$(p-2)$-brane,
stretched between the brane and its image. Such configurations are allowed
according to the rules of refs.\cite{9512059,9512062}. Naively, when the
D-$p$-brane and its
image coincide these stretched branes will have vanishing tension. But
quantum corrections must
give non-vanishing contribution to the tension,
reflecting the fact that these are non-BPS branes.\footnote{Otherwise
we
should expect a singularity in the moduli space of this system for
coincident D-brane $-$ orientifold plane system. This is known not to be
present.} Unfortunately
calculating tensions of these non-BPS branes in the weak coupling limit
remains an open problem.\footnote{As we shall see later, this problem is
related to finding the last row of table 1.} However as we shall see in
subsection \ref{s2}, for every
value of $p$ between 3 and 6, one can calculate the tensions of these
non-BPS branes in the strong coupling limit. The answer has been 
summarized in table 2.
\begin{center} 
\begin{tabular}{|c|c|c|} 
\hline $p$ & (tension)$^{1\over (p-2)}$ & $\wt C_p$
\\ 
\hline 6 & $\wt C_6 m_S$ & known \\ 
\hline 5 & $\wt C_5 m_S g_S^{-{1\over 6}}$
& known \\ 
\hline 4 & $\wt C_4 m_S g_S^{-{1\over 3}}$ & unknown \\ 
\hline 3 & $\wt C_3 m_S g_S^{-{1\over 2}}$ & known \\ 
\hline 
\end{tabular} 
\medskip

Table 2: Tensions of magnetically charged $(p-3)$-branes
on the D-$p$-brane 
O-$p$-plane system in the strong coupling limit.
\end{center}

The first column in this table describes the value of $p$ as before. The
second column represents the $(p-2)$-th root of the tension of the
magnetically charged $(p-3)$-brane. This root is taken in order to make it
into a quantity of mass dimension 1. $m_S$ and $g_S$ denote, as before,
the square root of the fundamental string tension and the string coupling
constant respectively, and $\wt C_p$ denote numerical constants. The last
column shows
that at present the coefficients $\wt C_p$ are known for $p=3$, 5 and 6,
but
not for $p=4$.

We again observe that there is a regularity in this spectrum. In
particular the $(p-2)$-th root of the tension of the $(p-3)$-brane on the
D-$p$-brane $-$ O-$p$-plane system goes as:
\be \label{e2}
m_S (g_S)^{{p\over 6}-1}\, .
\ee
Again it is natural to suspect that this reflects some deeper aspect of
string theory which is not understood at present.

\subsection{Strong coupling description of electrically charged states and
magnetically charged branes} 
\label{s2}

In this subsection we shall review the analysis leading to tables 1 and 2.
We
shall
discuss each value of $p$ separately, since the strong coupling
description of the D-$p$-plane $-$ O-$p$-plane system is different for
each value of $p$.

\noindent \underline{$p=6$}

In this case the system under study is a D6-brane on top of an
O6-plane in type IIA string theory. The strong coupling description of
this system is known to be M-theory on $R^{6,1}\times \NN$, where
$R^{6,1}$ is along the world-volume of the D6$-$O6 system, and $\NN$
is the double cover of the Atiyah-Hitchin space\cite{AH} with a rescaled
metric\cite{9607163,9803194}. Asymptotically,
$\NN$ locally looks like $R^3\times S^1$. The Planck mass $m_p$ of the
M-theory,
and the radius $R$ of this $S^1$ are related to $m_S$ and $g_S$ of type
IIA string theory via the relations:
\be \label{e3}
m_p=m_S (g_S)^{-{1\over 3}}, \qquad R = m_S^{-1} g_S\, .
\ee
The metric on $\NN$ is given by
\be \label{eathi}
ds^2 = {R^2\over 4} ds_{AH}^2\, ,
\ee
where $ds_{AH}^2$ is the standard Atiyah-Hitchin metric\cite{AH}.
The SO(2) gauge field $A$ on the brane world-volume is
related to the
three form gauge field $C_{\mu\nu\rho}$ of M-theory as
\be \label{ex1}
C = \omega \wedge A + \cdots
\ee
where $\omega$ is the unique normalizable harmonic two form on
$\NN$\cite{GIBRUB,MANSCH,9402032}, and $\cdots$ denotes terms
involving other normalizable and non-normalizable differential forms on
$\NN$.

The topology as well as the metric on $\NN$ is completely known. In
particular $\NN$ contains a non-contractible two cycle of minimal area
$-$ called the bolt $-$ which has the property that the integral of the
two form $\omega$ over the bolt is non-vanishing. 
{}From the relation
\refb{ex1} and the fact that a membrane is electrically 
charged under $C$,
it follows that a membrane wrapped on the bolt will be electrically
charged under $A$. In other words, 
the electrically charged stable non-BPS state on the world-volume of
the D6$-$O6
system is described by the M-theory membrane wrapped on the bolt of
$\NN$\cite{9803194}.
The area of the bolt is equal to $\pi^3 R^2$. On
the other hand, the membrane tension is proportional to $m_p^3$. Thus the
mass of the state is given by:
\be \label{e4}
C_6 m_p^3 R^2= C_6 m_S g_S\, ,
\ee
where $C_6$ is a known constant.

Following the same logic, the magnetically charged three brane on the
D6-O6 world-volume can be identified as the M-theory five-brane wrapped on
the bolt of $\NN$. The tension of this 3-brane can be calculated by
multiplying the five-brane tension ($m_p^6$) with the area of the bolt.
This is
given by
\be \label{ex2}
(\wt C_6)^4 m_p^6 R^2 = (\wt C_6)^4 (m_S)^4\, ,
\ee
where $\wt C_6$ is a known numerical constant.
Eqs.\refb{e4} and \refb{ex2} reproduce the first rows of tables 1 and 2
respectively.

\noindent \underline{$p=5$}

The system under study is a D5-brane on top of an O5-plane in type IIB
string theory. In the strong coupling limit, this theory is S-dual to the
weakly coupled type IIB string theory on $R^{5,1}\times
(R^4/(-1)^{F_L}\cdot\II_4)$ where $R^{5,1}$ is along the D5-O5 world
volume, $\II_4$ changes the sign of the coordinates of $R^4$ $-$ the
directions transverse to the D5-O5 world-volume, 
$-$ and $(-1)^{F_L}$ changes
the sign of all the Ramond sector states on the left-moving sector of the
string world-sheet\cite{9604070}. This can be argued by noting that under
S-duality of type IIB string theory $\Omega$ gets transformed to
$(-1)^{F_L}$ and a D5-brane is transformed to an NS 5-brane. Thus naively
one would think that the dual system should correspond to the orbifold
described above together with an NS 5-brane. But upon examining the
spectrum of massless states originating in the twisted sector of the
orbifold theory one finds that they are already in one to one
correspondence with
the massless degrees of freedom living on the D5-O5 system. Thus there is
no need to add another NS five brane; in fact adding it will double the
number of massless degrees of freedom, and will describe the dual of a
system of two D5-branes on top of an O5-plane.

The relationship between the string
scale $\wt
m_S$ and the coupling constant $\wt g_S$ of this dual theory, and those of
the original theory is given by:
\be \label{e5}
\wt m_S = m_S (g_S)^{-{1\over 2}}, \qquad \wt g_S = (g_S)^{-1}\, .
\ee
The SO(2) gauge field on the D5-brane $-$ O5-plane world volume
corresponds to massless
vector fields originating in the twisted sector of this orbifold theory.
The state carrying electric charge under the SO(2) gauge field corresponds
to,
in this orbifold theory, the
non-BPS D0-brane of IIB placed on the orbifold
plane\cite{9805019,9806155,9808141}. This has mass
\be \label{e6}
C_5 \wt m_S (\wt g_S)^{-1} = C_5 m_S (g_S)^{1\over 2}\, ,
\ee
where $C_5$ is a known constant. Similarly the two brane carrying magnetic
charge under this SO(2) gauge field corresponds to a non-BPS D2-brane of
type IIB string theory, placed inside the orbifold fixed plane. Its
tension is given by:
\be \label{e7}
(\wt C_5)^3 \wt m_S^3 (\wt g_S)^{-1} = (\wt C_5)^3 (m_S)^3 (g_S)^{-{1\over
2}}\, ,
\ee
where $\wt C_5$ is another known constant. Eqs.\refb{e6} and \refb{e7}
reproduce the second rows of tables 1 and 2 respectively.

\noindent \underline{$p=4$}

The configuration under study is a D4-brane on top of an O4-plane in type
IIA string theory. The strong coupling limit of this theory is best
described as M-theory on $R^{4,1}\times S^1\times (R^5/\II_5\cdot\sigma)$,
together with a five-brane (and its image under $\II_5\cdot\sigma$) placed
at the origin of $R^5$ with its
world-volume extending along $R^{4,1}\times S^1$\cite{9803194}.
Here
$R^{4,1}$ is along the world-volume of the original D4-O4 system, $S^1$ is
a circle of
radius $R$ given in eq.\refb{e3}, $\II_5$ reverses the sign of the
coordinates of $R^5$
transverse to the brane world-volume, and $\sigma$ denotes the
transformation which changes the sign of the three form gauge field of
M-theory.  This can be
seen by noting that under the type IIA - (M-theory on $S^1$) duality,
$\Omega$ of type IIA is mapped to $\sigma$ of M-theory, and the four
brane of type IIA is mapped to a five brane of M-theory wrapped on $S^1$.
The Planck mass $m_p$ of M-theory is given in terms on $m_S$ and
$g_S$ as in eq.\refb{e3}.
\begin{figure}[|ht]
\newpage
\begin{center}
\leavevmode
\epsfbox{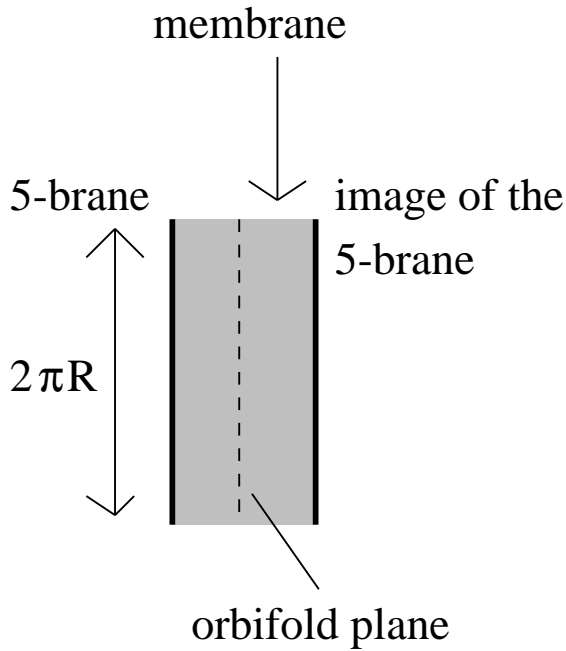}
\end{center}
\caption{Membrane stretched between the M5-brane and its image under
$\II_5\cdot\sigma$. We shall consider the case where the 5-brane (and
its image) coincide with the orbifold plane.} \label{f14}
\end{figure}

The five-brane world-volume carries a self-dual anti-symmetric tensor
field $B_{MN}$. The component $B_{1\mu}$, where $x^1$ denotes the
coordinate along $S^1$ and $\mu$ is the coordinate along $R^{4,1}$, is the
SO(2) gauge field $A_\mu$ on the D4-O4 system. As displayed in
Fig.\ref{f14}, the
world-volume of the five brane placed at the origin of $R^5$ also contains
a non-BPS string from the membrane stretched between the five-brane and
its image under $\II_5\cdot\sigma$\cite{9803194}.\footnote{Classically
this
string should
have zero tension when the five-brane approaches its image, 
but if this had been true also quantum mechanically
then the moduli space would have a singularity when the five brane
coincides with the orbifold plane. Using the duality between M-theory on
$T^5/Z_2$ and type IIB on 
K3\cite{9512196,9512219}, one can see that there is
no singularity
in this region of the moduli space.}  Although we cannot explicitly
compute the tension of this string, by dimensional analysis we see that
this tension must be proportional to $m_p^2$, since this is the only scale
in the problem.

Since the non-BPS string carries $B_{MN}$ charge, this string wrapped
on $S^1$ will be electrically charged under $B_{1\mu}=A_\mu$. The mass of
this state is given by:
\be \label{e8}
C_4 m_p^2 R = C_4 m_S (g_S)^{1\over 3}\, ,
\ee
where $C_4$ is an unknown numerical constant. On the other hand, the
non-BPS string with world-volume along $R^{4,1}$ will be magnetically
charged under the gauge field $B_{1\mu}=A_\mu$, and its tension will be
given by:
\be \label{e9}
(\wt C_4)^2 m_p^2 = (\wt C_4)^2 (m_S)^2 (g_S)^{-{2\over 3}}\, ,
\ee
where $\wt C_4$ is a numerical constant related to $C_4$. Eqs.\refb{e8}
and \refb{e9} reproduce the third rows of tables 1 and 2 respectively.

\noindent \underline{$p=3$}

The system under study is the D3-brane on top of an O3-plane in type IIB
string theory. The strong coupling limit of this system is dual to a
weakly coupled type IIB string theory in the same background, with the
parameters of the dual
theory related to those in the original theory by eq.\refb{e5}. The
electrically charged state in the original theory is mapped to the
magnetically charged state in the dual theory. Unfortunately at present we
do not know anything about this state, as was discussed earlier in
subsection
\ref{s1}. On the other hand, the magnetically charged state in the
original theory is mapped to the electrically charged state in the dual
theory. This is a perturbative open string state, and has mass
proportional to $\wt m_S$ for small $\wt g_S$. Thus the mass of the
magnetically charged state in the original theory in the strong coupling
limit is give by:
\be \label{e10}
\wt C_3 \wt m_S = \wt C_3 m_S (g_S)^{-{1\over 2}}\, ,
\ee
where $\wt C_3$ is a known numerical constant. This reproduces the last
row of table 2.

\subsection{Electrostatic self-energy of the electrically charged non-BPS
particle on the D3-brane $-$ O3-plane system} \label{s3}

In this subsection we shall give a lower bound on the electrostatic
self-energy of the electrically charged non-BPS particle on the D3-brane
$-$ O3-plane system in the strong coupling limit.
To do this we go to the dual weakly
coupled description where this particle corresponds to a magnetically
charged particle on the D3-brane $-$ O3-plane world-volume. Although we do
not know at present how to explicitly construct this state, it is clear
that sufficiently far away from the center, the magnetic field around the
state will look like the magnetic field of a point monopole. Let $r_c$
be
the distance beyond which this happens. If we normalize the gauge field on
the D3-brane so that the action has the form:
\be \label{ez1}
{1\over \wt g_S}\int d^4 x F_{\mu\nu} F^{\mu\nu}\, ,
\ee
where $\wt g_S$ as usual is the string coupling constant in this dual
string theory, then the magnetic field for $r>>r_c$ is of order $(1/r^2)$,
and hence its contribution to the total energy of the system from the
region $r\ge r_c$ is of order
\be \label{ez2}
{1\over \wt g_S}\int_{r\ge r_c} {d^3 r\over r^4} \sim (\wt
g_S)^{-1}(r_c)^{-1}\, .
\ee
In order to give a lower bound to this expression we need an upper bound
on $r_c$. This is obtained by noting that $r_c$ cannot be larger than the
string scale $(\wt m_S)^{-1}$ in this dual string theory, since for small
$\wt g_S$ we expect the lightest massive states in this theory to have
mass of order $\wt m_S$. Thus beyond the distance $(\wt m_S)^{-1}$, the
magnetic field of the monopole should approach that of a point
monopole.
This gives the following lower bound to the magnetostatic energy:
\be \label{ez3}
\wt m_S (\wt g_S)^{-1} \sim m_S (g_S)^{1\over 2}\, .
\ee
This exceeds the expected answer $m_S (g_S)^{1\over 4}$
from eq.\refb{e1}.

This suggests that eq.\refb{e1} is applicable, if at all, only to the
`intrinsic mass' of the non-BPS particle (if it could be defined at all),
and cannot account for the contribution from the Coulomb energy.
Presumably the issue will be clarified once we have an explicit
construction of this non-BPS state. It is the same problem which appears
in a more severe form in the case of $p=2$. Here the electrostatic
self-energy is infinite, and completely masks the `intrinsic mass' of the
particle.

\sectiono{Some Related Developments} \label{ss6}

In this section we shall briefly discuss some other related developments
in this field. In particular, we shall discuss
\begin{enumerate}
\item construction of
other non-BPS states in type I string
theory\cite{9810188}, 
\item relationship between D-brane charge and 
K-theory\cite{9810188,9812135}, and 
\item application of boundary state formalism to the study of non-BPS 
states\cite{9806155}.
\end{enumerate}
There are several other related
developments\cite{9901159,9902158,9903129,9812003,9807138,9804160,9808073,9902181} 
which will not be discussed here. At the end
we shall also briefly discuss some open
problems.

\subsection{Other non-BPS branes in type I string theory}

In the same way that we constructed a D0-brane in type I string theory,
one can construct a D8-brane in this theory. The idea is to start with the
non-BPS D8-brane of type IIB string theory, and mod it out by the
world-sheet parity transformation $\Omega$. The result is a non-BPS
D8-brane of type I string theory. The tachyonic mode of the open string
with both ends on the
8-brane is
projected out as in the case of the D0-brane. However in type I string 
theory
there are also space filling D9-branes, and it turns out that open
strings
with one end on the D8-brane and the other end on a D9-brane has
tachyonic modes which are not projected out\cite{9903123}. Thus these
branes are not stable.

One can also construct non-BPS D-instantons in the type I string theory as
follows\cite{9810188}. We can start from a D-instanton anti-D-instanton
pair of type IIB string theory, and mod out the theory by the world-sheet
parity transformation $\Omega$. The result is a non-BPS D-instanton of
type I string theory. One can show that the tachyonic mode is projected
out under this operation; so that the D-instanton is a stable
configuration of type I string theory. 

A similar construction can be done by starting with a D7-brane $-$
$\bd$7-brane pair of type IIB string theory, and modding out the
configuration by $\Omega$. Again the tachyonic mode originating in open
strings with both ends on the D7-brane is projected out. But in this case
there is a tachyonic mode in the open strings with one end on the D7-brane
and the other end on the D9-brane. Thus the D7-brane is not a stable
configuration in type I string theory\cite{9903123}.

\subsection{K-theory}

Another related development in this field has been the discovery of the
relationship between elements of the K-group of space-time manifold and
D-brane charges on the same manifold\cite{9810188}. This is related to
the idea of
representing a D-brane as a tachyonic soliton on a D-brane - anti-D-brane
pair of higher dimensions. The simplest example is that of type IIB string
theory, so we shall only discuss this case. In this case, following the
discussions of section \ref{ss4} we see that a BPS D-$(2p+1)$-brane can be
regarded as a soliton solution on a D-$(2p+3)$ $-$ $\bd$-$(2p+3)$-brane
pair. Each of the D-$(2p+3)$ branes on the other hand can be regarded as a
soliton solution on a D-$(2p+5)$-brane $\bd$-$(2p+5)$-brane pair.
Following
this argument we see that each stable D-brane in type IIB string theory
can be regarded as some kind of soliton solution on a sufficient number
($N$) of 9-brane $-$ anti-9-brane pair. It was shown by
Witten\cite{9810188} that
the solution representing a system of D-branes (possibly with some gauge
field configurations on them) can be completely classified by specifying
the $U(N)$ gauge bundles $E$ and $F$ on the 9-brane and the anti-9-brane
which characterize the gauge field configurations corresponding
to this soliton.
Furthermore, if we
add equal number of extra 9-branes and anti-9-branes to the system with
identical gauge bundles $H$ on them, then the tachyon associated with the
open strings stretched between the 9-brane and the anti-9-brane is a
section of a trivial bundle, and hence can condense to the minimum ($T_0$)
of the potential everywhere on the 9-brane $-$ anti-9-brane world-volume.
Since this configuration is identical to the vacuum, we conclude that
adding such extra pairs of 9-brane and anti-9-brane has no effect on the
topological class of the soliton.
Thus the D-brane charges are classified by specifying a pair of U(N)
vector bundles $(E,F)$ subject to the equivalence relation
\be \label{ek1}
(E,F)\equiv (E+H,F+H)\, ,
\ee
for any U(M) vector bundle $H$. This is precisely the definition of the
K-group of the space-time manifold.

This is the basic idea of using K-theory to classify D-brane charges.
Similar analysis can be carried out for type I and type IIA string
theories as well. In type I theory the starting point is the
representation of all D-branes as solitons on D9-$\bd$9-brane
system\cite{9810188},
whereas in type IIA string theory the starting point is the
representation of a D-brane as a soliton on a system of non-BPS
D9-branes\cite{9812135}.

\subsection{Boundary state approach to non-BPS branes}

The boundary state approach\cite{POLCAI,CLNY,ONOISH,ISHIBASHI,BOUNDARY} to
the study of
non-BPS
D-branes was
pioneered by Bergman and Gaberdiel\cite{9806155}. Corresponding to any
D-brane in string theory, we can associate a
boundary
state $|B\ra$ in the closed string sector whose inner product with a
closed string state describes the amplitude
for a closed string emission from the D-brane. Furthermore, if $|B\ra$ and
$|B'\ra$ denote the boundary states associated with a pair of (not
necessarily identical) D-branes, then $\la B|B'\ra$
describes the one loop partition function of an open string stretched from
the first D-brane to the second D-brane.  The boundary state $|B_p\ra$ of
a BPS D-$p$-brane in type IIA or type IIB string theory can be written as
a sum of two terms:
\be \label{ebou1}
|B_p\ra = {1\over\sqrt 2}(|NSNS_p\ra + |RR_p\ra)\, ,
\ee
where $|NSNS_p\ra$ and $|RR_p\ra$ denote the contribution to the boundary
state from the NSNS and RR sector closed strings respectively. The inner
product $\la B_p|B_q\ra$ can be expressed as
\be \label{ebou2}
\la B_p|B_q\ra = Tr_{p-q}{1 + (-1)^F\over 2} \, ,
\ee
where $Tr_{p-q}$ denotes trace over the open string states stretched from
the
D-$p$-brane to the D-$q$-brane. In this equation the contribution
proportional to 1 comes from the NSNS component of the boundary state,
whereas the contribution proportional to $(-1)^F$ comes from the RR
component of the boundary state.

In this notation the boundary state $|\wt B_p\ra$ describing the non-BPS
D-$p$-brane of
type IIB or type IIA string theory is given by\cite{9806155,9809111}
\be \label{ebou3}
|\wt B_p\ra = |NSNS_p\ra\, .
\ee
Note that the contribution from the RR sector is absent, reflecting the
fact that the non-BPS D-brane does not carry any RR charge. Also the NSNS
sector contribution to \refb{ebou3} has an extra factor of $\sqrt 2$
compared to that in \refb{ebou1}. This reflects the fact that the non-BPS
D-brane has an extra multiplicative factor of $\sqrt 2$ in its tension.

{}From eqs.\refb{ebou1}-\refb{ebou3} it follows that
\be \label{ebou4}
\la \wt B_p|\wt B_p \ra = Tr_{p-p}(1)\, .
\ee
This shows that the partition function of open string states living on
the non-BPS D-brane has no GSO projection.

One can also analyse the fate of stable non-BPS D-branes in various
orbifolds and orientifolds of type II string theories using the boundary
state approach. Let us consider, for example, the case of type I
D-particle. In this case the boundary state is described by
\be \label{ebou5}
{1\over \sqrt 2} (|\wt B_0\ra + 32 |B_9\ra + |C\ra)\, ,
\ee
where $|\wt B_0\ra$ is the boundary state of the non-BPS D0-brane (the
$(1/ \sqrt 2)$ factor is due to the $\Omega$ projection), $32|B_9\ra$
denotes the boundary state corresponding to the 32 BPS D9-branes in the
vacuum, and $|C\ra$ is the crosscap
state\cite{POLCAI,CLNY,ONOISH,ISHIBASHI} reflecting the
effect of $\Omega$ projection. The terms involving $|\wt B_0\ra$ 
in the inner product of
this boundary state with itself
is given by
\be \label{ebou6}
{1\over 2}(\la \wt B_0|\wt B_0\ra + \la \wt B_0|C\ra + \la C|\wt B_0\ra
+32 \la \wt B_0|\wt B_9\ra + 32
\la \wt B_9 | \wt B_0\ra)\, .
\ee
The sum of the first three terms gives
\be \label{ebou7}
Tr_{0-0}{1+\Omega\over 2}\, ,
\ee
where $Tr_{0-0}$ denotes trace over open strings with both ends on the
D0-brane. On the other hand the last two terms give
\be \label{ebou8}
32 Tr_{0-9}(1)\, ,
\ee
where $Tr_{0-9}$ denotes the trace over open string states stretched from
the D0-brane to the D9-brane. There is no $\Omega$ projection in this
term, since $\Omega$ relates these open strings to open strings stretched
from the D9-brane to the D0-brane. Thus the effect of $\Omega$ projection
is to simply include either the $0-9$ or the $9-0$ sector, but not both.

Since $|\wt B_0\ra$, $|C\ra$ and $|B_9\ra$ are all explicitly known, we
can evaluate each term in \refb{ebou6} explicitly. Comparing these with
\refb{ebou7} we can explicitly derive the $\Omega$ projection rules for
the open strings with both ends on the non-BPS D0-brane, and check that
these rules agree with the ones derived following the arguments in
subsection \ref{stypei}. In particular, one can verify that
the tachyonic mode on this
D-particle
is projected out under $\Omega$.

\subsection{Open questions and speculations} 

We shall conclude this article by reviewing some of the open questions
and with some speculations.
\begin{enumerate}
\item The various arguments given in favour of the idea that the tachyonic
ground state on the brane anti-brane pair is indistinguishible from the
vacuum are all indirect, and involves first compactifying one or more 
directions
tangential to the brane world-volume, followed by switching on the
tachyon vev and then taking the radius back to infinity. A direct proof of
this on a non-compact brane-antibrane pair, presumably based on the
construction of an explicit classical solution in the open string field
theory on the brane-antibrane pair describing the tachyonic ground state,
is still lacking. Similarly, one should be able to construct an explicit
classical solution in this open string field theory representing the
tachyonic kink solution and show that this solution describes a non-BPS
D-brane.

\item One of the difficulties in understanding the phenomenon of tachyon
condensation on the brane-antibrane pair has been in understanding what
happens to the various U(1) gauge fields living on the original system.
The tachyon is charged under one combination of the two U(1) gauge fields,
and hence breaks this gauge symmetry. However the other linear
combination, which we shall denote by $\AAA_\mu$, does not get broken
since the tachyon, as well as all other perturbative open string states
living on the brane-antibrane world-volume, are neutral under this gauge
field.

It has been suggested in ref.\cite{9901159} that the other U(1) is in the
confining phase. The suggested mechanism for this confinement is
the condensation of the tachyonic $(p-3)$-branes obtained from
D-$(p-2)$-branes stretched between the original D-$p$-brane and the
anti-D-$p$-brane. Thus for example for $p=3$, it involves condensation of
the tachyonic mode of the D-string stretched between the D3-brane and the
$\bd$3-brane. It was shown in \cite{9901159} that this tachyon is
magnetically charged under $\AAA_\mu$, and hence condensation of this
tachyon will imply that the corresponding U(1) gauge theory is in the
confining phase.

Whereas the general idea is quite appealing, this mechanism is highly
non-perturbative from the point of view of the world-volume theory of the
D3-brane $-$ $\bd$3-brane pair. On the other hand the indirect arguments
reviewed
in this article showing that the tachyonic ground state is identical to
the vacuum configuration are based on open string tree level analysis.
Thus there must be a way to see the phenomenon of confinement of the
U(1) gauge field $\AAA_\mu$ at open string tree level. Presumably once we
understand how to
describe tachyon condensation using classical open string field theory,
this issue will be automatically resolved.

\item Another open problem, which has already been discussed earlier, is
the construction of magnetically charged non-BPS D-$(p-3)$-brane on the
D-$p$-brane $-$ O-$p$-plane system. It is clear that these stable branes
must exist in the spectrum, so one should be able to find them in the
weakly coupled string theory.

\item It would be interesting to investigate the relationship
between weak coupling perturbation expansion for the mass of a non-BPS
state and its strong coupling limit. This might give us new insight into
string theory at finite coupling.

\item One of the main lessons from our analysis (and of 
refs.\cite{9510227,9704006,9708075})
is that the existence of tachyons in the spectrum of open
string theory does
not necessarily signify a sickness of the theory, but often simply
indicates the
existence of a ground state with energy (density) lower than
that of the starting configuration. It would be interesting to 
investigate
if closed string tachyons have a similar interpretation.

\item {}From our discussion in this article it is clear that all D-branes
in type IIA (IIB) string theory can be regarded as classical solutions in
the open string field theory living on a system of non-BPS D9-branes
(D9-brane $-$ $\bd$9-brane pair). It would be interesting to see if this
can also be done for other known objects in string theory, namely the
fundamental string and the NS 5-branes.\footnote{K-theory does not contain
these states, but K-theory uses only a small subset of available open
string fields, namely the tachyon and the gauge bosons.} Actually
fundamental
strings appear as bound state poles in the S-matrix computed from Witten's
open string field
theory\cite{WITTBFT,WITTSFT,CUBIC}. On the other hand a formal
construction was presented in \cite{CUBIC} showing that any string
background represented by a two dimensional conformal field theory (of
which the NS five-brane is an example) can be
represented as a classical solution in the purely cubic open string field
theory. If these results can be made more concrete, then one could take
open
string field theory on the non-BPS D9-brane (D9-$\bd$9 brane pairs) as the
fundamental formulation of type II string
theories and their orbifolds/orientifolds, since all states in string
theory could be constructed from this field theory. 

\end{enumerate}


\noindent{\bf Acknowledgement}: I would like to thank
O.~Bergman, S.~Elitzur, M.~Gaberdiel, P.~Horava, 
N.~Manton, B.~Pioline, E. Rabinovici, A.~Recknagel,
V.~Schomerus and E. Witten 
for useful correspondence at various stages of this work.
I would also like to
thank the organisers of the APCTP winter school for organising an
excellent school, and the Tata Institute at Mumbai,  IAS at the Hebrew
University and CTP at MIT 
for their hospitality during the preparation of
this manuscript.

\appendix

\sectiono{Conformal Field Theory of the Tachyonic Kink Solution}
\label{ssa}

From our discussion in sections \ref{ss2} and \ref{ss4}, it follows that a
non-BPS
D-$2p$-brane of type IIB string theory has two descriptions:
\begin{itemize}
\item D-$2p-\bd$-$2p$ of type
IIA string theory modded out by $(-1)^{F_L}$, and
\item tachyonic kink on D-$(2p+1)-\bd$-$(2p+1)$ system of type IIB string
theory.
\end{itemize}
In this appendix we shall address the issue of
proving the equivalence of these two descriptions.
We shall focus on the non-BPS D0-brane of IIB, but extension to
the general case (non-BPS D-$2p$-brane of IIB and non-BPS D-$(2p+1)$-brane
of IIA) is straightforward. The details of the analysis of this
appendix can be found in ref.\cite{9808141}.
Some related  analysis for bosonic string
theory can be found in 
refs.\cite{9402113,9404008,9811237,9902105,9807161}. 

The outline of the proof is as follows. We begin with the
observation
that the tachyonic kink on the D1$-\bd$1 pair is
a classical solution in the open string field theory living on
the D1$-\bd$1 pair. Thus 
this configuration must be describable by a two
dimensional boundary conformal field theory.
Hence we need to
\begin{itemize}
\item find this CFT, and
\item show that this is equivalent to the CFT describing
$ {\rm D}0-\bd0 \hbox{ of IIA} / (-1)^{F_L}$.
\end{itemize}
The next question is: how do we find the CFT describing the kink?
This is done using the following steps.
\begin{itemize}
\item Find a series of marginal deformations which connect the
D1$-\bd$1 pair to the tachyonic kink.
\item Follow what happens to the CFT describing the D1$-\bd$1
pair under this marginal deformation.
\end{itemize}

Thus
our first job will be to find this series of marginal
deformations.
This is done in several steps.
\begin{enumerate}
\item
Compactify one direction along a circle $S^1$ of
radius $R$ and take the D1-$\bd$1 pair to lie along $S^1$.
Let
$x$ be the coordinate along $S^1$, and
$A_\mu$, $\bar A_\mu$ be the U(1) gauge fields on D1, $\bd$1-branes
respectively.
The first step is to
{\it
increase $\bar A_x$ from 0 to $1/2R$.} This is a marginal deformation
using boundary operators, and gives
\be \label{ea1}
\exp(i\ointop \bar A_x dx) = -1\, .
\ee
In the presence of such a Wilson line,
open strings with CP factors
$I$ and $\sigma_3$ are periodic  under $x\to x+2\pi R$, since they are
neutral under
$\bar A_x$, whereas open strings with CP factors
$\sigma_1$ and $\sigma_2$ are anti-periodic under $x\to x+2\pi R$ since
they carry unit
charge under $\bar A_x$.

\item Let $T$ denote the tachyon field originating in the
sector $\sigma_1$. This has a Fourier expansion of the form:
\be \label{ea2}
T(x,t) = \sum_{n\in Z} T_{n+{1\over 2}}(t) e^{i(n+{1\over
2}){x\over R}}\, ,
\ee
since it is anti-periodic under $x\to x+2\pi R$.
The
mass of the mode $T_{n+{1\over 2}}$ is given by
\be \label{ea3}
m^2_{n+{1\over 2}} = {(n+{1\over 2})^2\over R^2}-{1\over 2}\, .
\ee
We now note that
\begin{itemize}
\item For $R\le (1/\sqrt 2)$ there are no tachyonic modes.
\item For $R=1/\sqrt 2$, $T_{\pm{1\over 2}}$ is massless and
hence represent marginal boundary operators in the CFT.
\end{itemize}
In this second step 
{\it
we reduce $R$ from its initial value down to $1/\sqrt 2$.} This
corresponds to a marginal deformation involving bulk operators.

\item
As we shall see later, at $R={1\over \sqrt 2}$, $(T_{1\over
2}-T_{-{1\over 2}})$ corresponds to an exactly marginal operator.
In this third step
{\it we switch on vacuum expectation value (vev) of $(T_{1\over
2}-T_{-{1\over 2}})$.} This is a
marginal deformation involving a boundary operator.
\begin{figure}[|ht]
\begin{center}
\leavevmode
\epsfbox{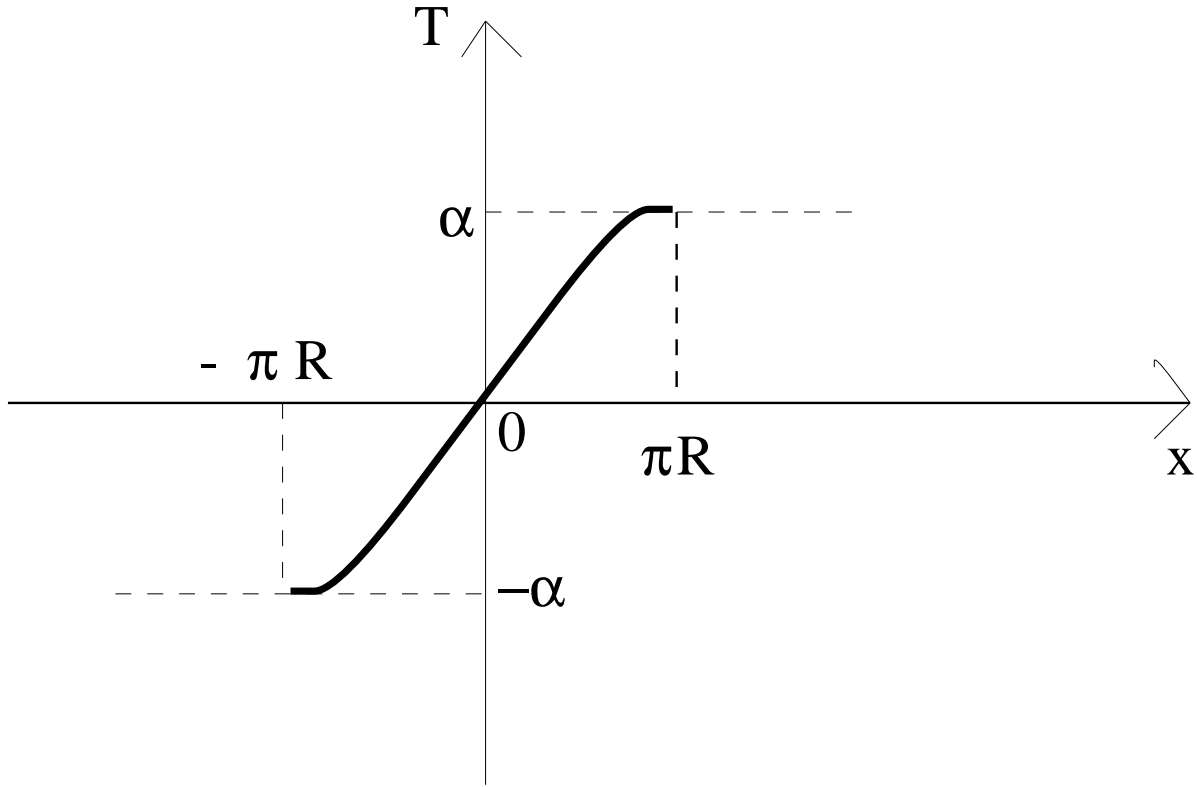}
\end{center}
\caption{Effect of switching on vacuum expectation value of $(T_{1\over 2}
-T_{-{1\over 2}})$.} \label{f9}
\end{figure}

The physical interpretation of switching on the vev of $(T_{1\over 2}
-T_{-{1\over
2}})$ is as follows.
If we take:
\be \label{ea3a}
(T_{1\over 2}-T_{-{1\over 2}})=-i\alpha, \qquad
T_{1\over 2}+T_{-{1\over 2}}=0, \qquad T_n=0 \quad
\hbox{for}\quad |n|>{1\over 2},\, ,
\ee
then 
\be \label{ea4}
T(x)=\alpha \sin{x\over 2R}\, .
\ee
As shown in Fig.\ref{f9}, this represents a kink.

Note that we have not said so far how much vev we should give to
$(T_{1\over 2}-T_{-{1\over 2}})$. This will be discussed shortly.

\item
{\it
After switching on the tachyon vev, we take the radius back to
infinity.} This corresponds to marginal deformation by a bulk operator.
\begin{figure}[|ht]
\begin{center}
\leavevmode
\epsfbox{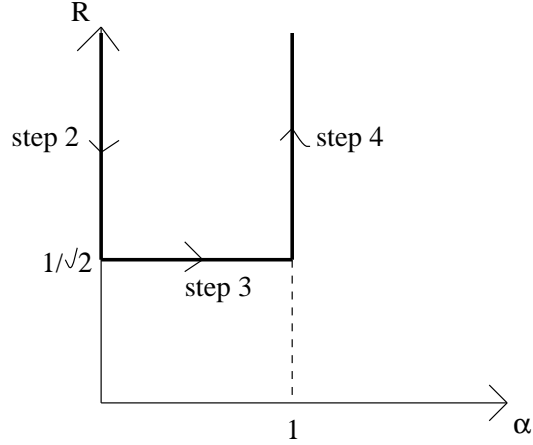}
\end{center}
\caption{The flow in the $R-\alpha$ plane from $(R=\infty,\alpha=0)$ to
$(R=\infty,\alpha=1)$.} \label{f10}
\end{figure}

It will be shown later that when we switch on this marginal deformation, 
for $R>{1\over \sqrt 2}$, $(T_{1\over 2}-T_{-{1\over
2}})$ develops a tadpole for a generic $\alpha$.
This is 
not surprising, since
for $R>{1\over \sqrt 2}$, $(T_{1\over 2}-T_{-{1\over
2}})=\alpha$ is not expected to be a solution of the equations of
motion for arbitrary $\alpha$.
However, we find that there are two values of $\alpha$ for which
the tadpole vanishes: namely
$\alpha=0$ and $\alpha=1$ (with a suitable normalization of the tachyon
field).
If we take the limit $R\to\infty$ at $\alpha=0$, we get back the
D1-$\bd$1 pair.
But if we take the $R\to\infty$ limit at $\alpha=1$, we should expect
to get the
kink on the D1-$\bd$1 pair.

This analysis also determines the amount of vev of $(T_{1\over 2} -
T_{-{1\over
2}})$ to be switched on at the third step. Namely, it should correspond to
$\alpha=1$.
\end{enumerate}
The steps 2, 3 and 4 correspond to the flow in the $(R,\alpha)$ plane as
shown in
Fig.\ref{f10}. Note that if we want to go from the
$(R=\infty,\alpha=0)$ point to the $(R=\infty,\alpha=1)$ point
directly, we need to perturb by $(T_{1\over 2} - T_{-{1\over 2}})$ at
$R=\infty$, which is a relevant boundary operator.

We now carry out these steps in detail and see what conformal field
theory
we get at the end of these steps. Since
marginal deformations up to the end of step 2 is straightforward, 
we focus on steps 3 and 4.
Thus our starting point will be the CFT at the end of step 2. This
corresponds to
\be \label{ea5}
R={1\over \sqrt 2}, \quad \exp(i\ointop\bar A_x dx)=-1, \quad
\alpha=0\, .
\ee
The
relevant world-sheet fields are a scalar field 
$X\equiv(X_L+X_R)$ representing the coordinate along $S^1$, and a
Majorana fermion $(\psi_L,\psi_R)$
representing the
world-sheet supersymmetric partner of $X$.
We impose Neumann boundary condition on $X$ and $\psi$:
\be \label{ea6}
X_L=X_R\equiv X_B/2, \qquad \psi_L=\psi_R\equiv \psi_B\, ,
\ee
where the subscript $B$ stands for boundary value.\footnote{For
simplicity we shall concentrate on the NS sector states throughout this
section, but a similar analysis can be carried out for the R-sector
states as well.}
Besides these fields, there are nine other bosonic coordinate
fields, their fermionic partners, and ghost fields, but these will not
play any crucial role in our analysis.

Let us now define $h$ to be the transformation $X\to X+2\pi R$.
The $h$ and $(-1)^F$ quantum numbers of the open string states carrying
different CP factors are then given as in table 3.
\begin{center} 
\begin{tabular}{|c|c|c|}
\hline
CP factor & $h$ & $(-1)^F$ \\
\hline
$I$, $\sigma_3$ & 1 & 1 \\
\hline
$\sigma_1$, $\sigma_2$ & -1 & -1 \\
\hline
\end{tabular}

\medskip

Table 3: The $(-1)^F$ and $h$ quantum numbers of various open string
states at the end of step 2.
\end{center}

Using these rules we can determine the complete spectrum of open
strings. In particular
vertex operator for $(T_{1/2}-T_{-(1/2)})$ in the 0-picture\cite{FMS} is
given by:
\be \label{ea7}
V_T \sim \psi_B (e^{iX_B/\sqrt 2}+e^{-iX_B/\sqrt
2})\otimes
\sigma_1\, .
\ee
This is odd under $h$ and $(-1)^F$.

We now use the fact that
at $R=(1/\sqrt 2)$, a free boson $X$ is equivalent to a pair
of Majorana fermions $(\xi,\eta)$. The relationship is of the form:
\be \label{ea8}
e^{i\sqrt 2X_L}\sim (\xi_L+i\eta_L), \qquad
e^{i\sqrt 2X_R}\sim (\xi_R+i\eta_R)\, .
\ee
Thus we have  three Majorana fermions
$\xi$, $\eta$, and $\psi$.
We can now rebosonize them as follows:
\be \label{ea9}
e^{i\sqrt 2\phi_{L\atop R}}\sim (\xi_{L\atop R}+i\psi_{L\atop R})\, ,
\ee
or
\be \label{ea10}
e^{i\sqrt 2\phi'_{L\atop R}}\sim (\eta_{L\atop R}+i\psi_{L\atop
R})\, .
\ee
$\phi$ and $\phi'$ are scalar fields.
The relationship between the currents in the bosonic and the fermionic
variables are as follows:
\ben \label{ea11}
&& \xi_L\eta_L\sim \p X_L, \qquad \xi_L\psi_L\sim \p \phi_L,
\qquad \eta_L\psi_L\sim \p\phi'_L\, , \nonumber \\
&& \xi_R\eta_R\sim \p X_R, \qquad \xi_R\psi_R\sim \p \phi_R,
\qquad \eta_R\psi_R\sim \p\phi'_R\, .
\een
{}From eq.\refb{ea6} and \refb{ea8}-\refb{ea11} we can easily see that
putting
Neumann boundary condition on $X$ and $\psi$ corresponds to
putting Neumann boundary condition on $\phi,\phi',\xi$ and $\eta$:
\ben \label{ebounx}
&& \phi_L=\phi_R\equiv{1\over 2}\phi_B, \qquad
\phi'_L=\phi'_R\equiv{1\over 2}\phi_B\, , \nonumber \\
&& \xi_L=\xi_R\equiv \xi_B, \qquad \eta_L=\eta_R\equiv \eta_B\, .
\een

We can now rewrite the vertex operator for the tachyon field in
terms of the new fields:
\begin{eqnarray} \label{ea11a}
V_T &\sim& \psi_B (e^{iX_B/\sqrt 2}+e^{-iX_B/\sqrt
2})\otimes\sigma_1 \nonumber \\
&\sim& \psi_B\xi_B \otimes \sigma_1 \sim \p\phi_B \otimes
\sigma_1 \, .
\end{eqnarray}
Now,
$\p\phi_B\otimes \sigma_1$ can be interpreted as the vertex operator of a
constant gauge
field $\AAA_\phi$ along $\phi$.
Hence it corresponds
to an exactly marginal deformation, as claimed earlier. 
Furthermore,
$\AAA_\phi$ is a
periodic variable. Let us denote by $\alpha$ a suitably normalized
$\AAA_\phi$ such that $\alpha$ has periodicity 2.

We shall now study the effect of switching on $\AAA_\phi$ on the
open string spectrum.
This can be done as follows:
\begin{itemize}
\item First of all, since $I$ and $\sigma_1$ commute with $\sigma_1$, we
conclude that
open string states with CP factors $I$, $\sigma_1$ are
neutral under $\AAA_\phi$.
Thus the spectrum in these sectors remain unchanged.

\item Since
\be \label{ea13}
[\sigma_1,\sigma_3\pm i\sigma_2]=\mp 2(\sigma_3\pm
i\sigma_2)\, ,
\ee
we see that
open strings in sectors $\sigma_3\pm i\sigma_2$ carry equal and
opposite
charges under $\AAA_\phi$.
Thus in these sectors switching on $\AAA_\phi$ causes a shift in the
$\phi$ momentum quantization rule:
\be \label{ea14}
p_\phi\to p_\phi\pm {\alpha\over \sqrt 2}\, .
\ee
The coefficient of $\alpha$ in this equation has been fixed by
requiring that $\alpha$ has periodicity 2. {}From
eqs.\refb{ea8}-\refb{ea10}
we see that under $(-1)^F\cdot h$, $\phi\to\phi + \sqrt 2\pi$ and
$\eta\to -\eta$. Thus projection under $(-1)^F\cdot h$ requires that in
each CP sector,
for fixed set of $\eta$ oscillators, $p_\phi$ is quantized as $n\sqrt 2$ +
a constant additive term, where $n$ is an integer. {}From this we see
that shifting
$\alpha$ by 2 does not change the quantization law of $p_\phi$.
\end{itemize}

Using these rules we can find the spectrum of open strings for
all values of $\alpha$, including at $\alpha=1$.
It turns out that the net result for the spectrum at $\alpha=1$ is that
in the sectors $(\sigma_3\pm i\sigma_2)$ the GSO projection gets
reversed, without any change in the $h$-projection.
Thus the $(-1)^F$ and $h$ quantum numbers carried by various open string
states at $\alpha=1$ are as given in table 4.
\begin{center}
\begin{tabular}{|c|c|c|}
\hline
CP factor & $h$ & $(-1)^F$ \\
\hline
$I$ & 1 & 1 \\
\hline
$\sigma_1$ & -1 & -1 \\
\hline
$\sigma_2$ & -1 & 1 \\
\hline
$\sigma_3$ & 1 & -1 \\
\hline
\end{tabular}

\medskip

Table 4. The $(-1)^F$ and $h$ quantum numbers of open string states at
$\alpha=1$.
\end{center}

\begin{figure}[|ht]
\begin{center}
\leavevmode
\epsfbox{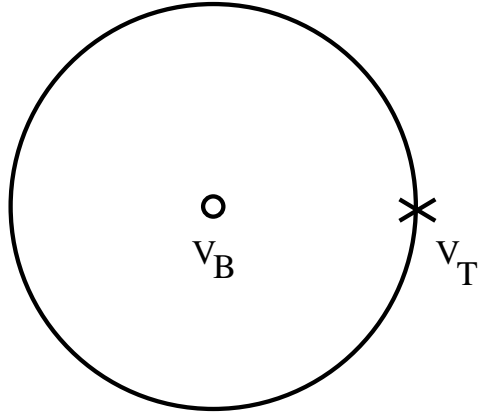}
\end{center}
\caption{Diagram contributing to tachyon one point function to first order
in $\delta R$.
At the boundary of the disk we must insert a factor of
$\exp(i(\alpha/ 2^{3/2})\sigma_1 \ointop \p \phi_B)$.
} \label{f11}
\end{figure}

This concludes the analysis in step 3. Note that when we combine the
spectrum from all the sectors, there is no net $h$ or $(-1)^F$
projection, since all combinations of these quantum numbers appear in
the spectrum. Thus we can use all combinations of $\xi$, $\eta$ and
$\psi$ oscillators to create a state from the Fock vacuum. If we use
$\phi'$ and $\xi$ as independent variables, then from \refb{ea10} we
see
that all the states are invariant under $\phi'_L\to \phi'_L+\sqrt
2\pi$, $\phi'_R\to \phi'_R+\sqrt
2\pi$. Since $\phi'=\phi'_L+\phi'_R$, this effectively corresponds to
$\phi'$ having a radius $\sqrt 2$. This fact will be useful to us
later.

We now proceed to step 4.
This involves
switching on the radius deformation and taking the $R\to\infty$ limit.
The
computation of a correlation function of open string vertex
operators on a disk for a generic value of $R$ and $\alpha$ involves
\begin{itemize}
\item inserting $\exp({i\alpha\over 2\sqrt
2}\sigma_1\ointop\p\phi_B)$ at the boundary, corresponding to the
$\alpha$-deformation,\footnote{The coefficient of $\alpha$ in the
exponent has again been fixed by demanding that shifting $\alpha$ by 2
does not change the S-matrix except for a redefinition of the external
open string states.}
\item inserting  $\exp(C\int d^2z \p X_L\p X_R)$ in the interior of the
disk corresponding to the radius deformation away from $R={1\over \sqrt
2}$,
\item inserting open string vertex operators corresponding to
external states on the boundary, and
\item inserting appropriate number of picture changing operators.
\end{itemize}
As an example we have displayed in Fig.\ref{f11} the diagram relevant for 
the computation of tachyon one point function to first order
in $\delta R\equiv (R-{1\over \sqrt 2})$.
Here $V_B\sim e^{-\Phi_L-\Phi_R}\psi_L\psi_R$ is the closed string vertex
operator in the $(-1,-1)$ picture representing radius deformation,
$(\Phi_L,\Phi_R)$ are the left- and the right-moving components of the
bosonized ghosts\cite{FMS}, and
$V_T\sim \p\phi_B\otimes \sigma_1$ is the tachyon vertex operator in the
$(0,0)$ picture.
This diagram can be easily computed, and the
final result is that:
\be \label{ea15}
\langle V_T\rangle \propto \sin(\alpha\pi)\, .
\ee
This vanishes at $\alpha=0,1$. As mentioned earlier, the $\alpha=0$ point
corresponds to the original D-string anti-D-string pair, whereas the
$\alpha=1$ point corresponds to the tachyonic kink solution on this pair.
\begin{figure}[|ht]
\begin{center}
\leavevmode
\epsfbox{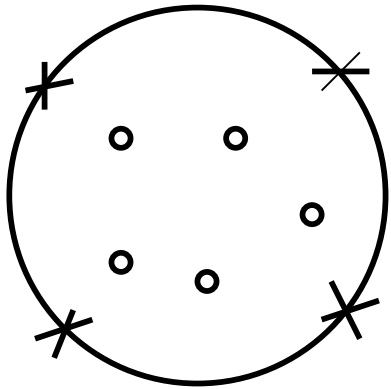}
\end{center}
\caption{Diagrams contributing to the open string tree level amplitude at
a general value of
$R$ and at $\alpha=1$. The circles denote closed string vertex operators 
$V_B$ corresponding to the radius deformation, and the crosses denote the
open string vertex operators corresponding to external open string states. 
At the boundary of the disk, there is also
an insertion of the operator 
$\exp(i(1/ 2^{3/2})\sigma_1 \ointop \p \phi_B)$.
} \label{f12}
\end{figure}

We shall from now on focus on the point $\alpha=1$, and analyse the system
at a general value of $R$.
For general $R$, and a general correlation function of open
string vertex operators, we need to sum over all possible number of
insertions of $V_B=\p X_L\p X_R$, representing the radius deformation
in
the (0,0) picture, in the interior of the disk with appropriate weight
factors. Since we have set $\alpha=1$, we have to insert a factor of
\be \label{ea16}
\exp\Big(i{1\over 2\sqrt 2}\sigma_1 \ointop \p
\phi_B\Big)\, ,
\ee
at the boundary of the disk. The effect of this insertion can be analysed
by shrinking
the contour integral along the boundary to inside the
disk, and picking up the residue at each insertion of $V_B$. The final
result is that\cite{9808141} it converts each $\p X_L\p
X_R$ to
$-\p\phi'_L\p\phi'_R$. Such a perturbation corresponds to decreasing the
$\phi'$ radius. 
Thus we conclude that increasing the $X$ radius at $\alpha=1$ is
equivalent to
decreasing the $\phi'$ radius at $\alpha=0$.
In particular,
the limit $R_X\to\infty$ gets converted to $R_{\phi'}\to 0$.

If we
introduce a new T-dual variable $\phi''$:
\be \label{ea17}
\phi''_L=\phi'_L, \qquad \phi''_R=-\phi'_R, \qquad
R_{\phi''}=1/R_{\phi'}\, ,
\ee
then
as $R_{\phi'}\to 0$, $R_{\phi''}\to\infty$.
At the same time,
Neumann boundary condition on $\phi'$ 
corresponds to
Dirichlet boundary condition on $\phi''$
\be \label{ea18}
\phi''_L=-\phi''_R \quad \hbox{at the boundary}\, .
\ee
The net result is that we have
a non-compact bosonic coordinate $\phi''$ with
Dirichlet boundary condition along $\phi''$.

Thus we conclude that the tachyonic kink solution on a D1+$\bd$1-brane 
corresponds to
a D0-brane at $\phi''=0$, where $\phi''$ is a new non-compact bosonic
coordinate. 
We can compute
the spectrum of open strings with ends on this D0-brane
by starting with the known spectrum at the end of
step 3 as given in table 4, 
and following it adiabatically as $R_{\phi''}$ increases. We saw that
at the end of step 3, the combined spectrum from all sectors has no
projection, and $\phi'$ behaves like a bosonic coordinate of radius
$\sqrt 2$. Thus $\phi''$ has radius $(1/\sqrt 2)$, $-$ the same as that
of $X$. As we increase the $X$ radius to some arbitrary value $R$,
$\phi''$ radius also gets increased to $R$. Thus the combined spectrum
of open strings will be that on a D0-brane on a circle of radius $R$,
with no GSO projection. This is identical to the one obtained by
modding out
the D0-$\bd$0 pair of IIA on a circle of radius $R$ by $(-1)^{F_L}$, as
studied in section
\ref{ss2}.

This shows that a tachyonic kink on the D-string
anti-D-string pair of type IIB string theory corresponds to a non-BPS
D0-brane of type IIB string theory as defined in section \ref{ss2}. The
fact that the non-compact coordinate is called
$\phi''$ and not $X$ is not of any relevance; all that matters is the CFT
describing the system and not how we label the CFT. 

As stated at the beginning of this appendix, 
similar analysis can be done for showing the equivalence of the
tachyonic kink on the D-$p-\bd$-$p$ pair on IIA (IIB), and
$[{\rm D}-(p-1)-\bd-(p-1)] /(-1)^{F_L}$ in IIB (IIA). Here
$p$ is an even integer in type IIA string theory, and is an
odd integer in type IIB string theory.

One can also consider a T-dual version of the analysis described here to
interpolate between a
a D0-brane $-$
$\bd0$-brane pair situated at diametrically opposite points on a circle,
and a
non-BPS D-string wrapped on the same circle. Running the flow backwards,
we see that there is a series of marginal deformations which take us from
a non-BPS D-string in type IIA string theory to a D0-brane $-$
$\bd0$-brane pair in the same theory. By analysing what background
corresponds to this deformation on the non-BPS D-string, one discovers
that it describes a kink-antikink pair\cite{9812031}. This allows us to
identify a
tachyonic kink on the non-BPS D-string to a BPS D-particle of type IIA
string theory. This can easily be generalized to show that a
tachyonic kink on
the non-BPS D-$(p-1)$ brane
corresponds to
the BPS D-$(p-2)$ brane in the same theory. Again $p$ is even for
type IIA
string theory and odd for type IIB string theory.


\begin{thebibliography}{99}

\bibitem{9803194}
A.~Sen,
JHEP {\bf 06}, 007 (1998)
hep-th/9803194.

\bibitem{9805019}
A.~Sen,
JHEP {\bf 08}, 010 (1998)
hep-th/9805019.

\bibitem{9805170}
A.~Sen,
JHEP {\bf 08}, 012 (1998)
hep-th/9805170.

\bibitem{9806155}
O.~Bergman and M.R.~Gaberdiel,
Phys. Lett. {\bf B441}, 133 (1998)
hep-th/9806155.

\bibitem{9808141}
A.~Sen,
JHEP {\bf 09}, 023 (1998)
hep-th/9808141.

\bibitem{9809111}
A.~Sen,
JHEP {\bf 10}, 021 (1998)
hep-th/9809111.

\bibitem{9810188}
E.~Witten,
JHEP {\bf 12}, 019 (1998)
hep-th/9810188.

\bibitem{9812031}
A.~Sen,
JHEP {\bf 12}, 021 (1998)
hep-th/9812031.

\bibitem{9812135}
P.~Horava,
hep-th/9812135.

\bibitem{9901014}
O.~Bergman and M.R.~Gaberdiel,
JHEP {\bf 03}, 013 (1999)
hep-th/9901014.

\bibitem{9812226}
H.~Garcia-Compean,
hep-th/9812226.

\bibitem{9901042}
S.~Gukov,
hep-th/9901042.

\bibitem{9902102}
K.~Hori,
hep-th/9902102.

\bibitem{9902116}
E.~Sharpe,
hep-th/9902116.

\bibitem{9902160}
O.~Bergman, E.G.~Gimon and P.~Horava,
hep-th/9902160.

\bibitem{9904153}
K.~Olsen and R.J.~Szabo,
hep-th/9904153.

\bibitem{DBRANE}
J. Dai, R. Leigh, and J. Polchinski, Mod. Phys. Lett. 
{\bf A4} (1989) 2073; \\
R. Leigh, Mod. Phys. Lett. {\bf A4} (1989) 2767; \\
J. Polchinski, Phys. Rev. {\bf D50} (1994) 6041 [hep-th/9407031]; \\
J. Polchinski, S. Chaudhury and C. Johnson, [hep-th/9602052]; \\
J. Polchinski, [hep-th/9611050].

\bibitem{9403040}
M.B.~Green,
Phys. Lett. {\bf B329}, 435 (1994)
hep-th/9403040.

\bibitem{9511194}
T.~Banks and L.~Susskind,
hep-th/9511194.


\bibitem{9604091}
M.B.~Green and M.~Gutperle,
Nucl. Phys. {\bf B476}, 484 (1996)
hep-th/9604091.

\bibitem{9604156}
G.~Lifschytz,
Phys. Lett. {\bf B388}, 720 (1996)
hep-th/9604156.

\bibitem{9612215}
V.~Periwal,
hep-th/9612215.

\bibitem{SCHPHYSREP}
J.H.~Schwarz,
Phys. Rept. {\bf 89}, 223 (1982).

\bibitem{9503124}
E.~Witten,
Nucl. Phys. {\bf B443}, 85 (1995)
hep-th/9503124.

\bibitem{9506160}
A.~Dabholkar,
Phys. Lett. {\bf B357}, 307 (1995)
hep-th/9506160.

\bibitem{9506194}
C.M.~Hull,
Phys. Lett. {\bf B357}, 545 (1995)
hep-th/9506194.

\bibitem{9510169}
J.~Polchinski and E.~Witten,
Nucl. Phys. {\bf B460}, 525 (1996)
hep-th/9510169.

\bibitem{9603167}
M.R.~Douglas and G.~Moore,
hep-th/9603167.

\bibitem{9507158}
K. Becker, M. Becker and A. Strominger, Nucl. Phys. {\bf B456} 130 (1995) 
hep-th/9507158.

\bibitem{9507012}
P. Aspinwall, Phys. Lett. {\bf B357} 329 (1995) hep-th/9507012.

\bibitem{ORIENT}
A. Sagnotti, `Open Strings and their Symmetry Groups', Talk at
Cargese Summer Inst., 1987; \\
G. Pradisi and A. Sagnotti, Phys. Lett. {\bf B216} 59 (1989); \\
M. Bianchi, G. Pradisi and A. Sagnotti, Nucl. Phys. {\bf B376}
365 (1992); \\
P. Horava, Nucl. Phys. {\bf B327} 461 (1989),
Phys. Lett. {\bf B231} 251 (1989).

\bibitem{9601038}
E. Gimon and J. Polchinski, Phys. Rev. {\bf D54} 1667 (1996) 
[hep-th/9601038]. 

\bibitem{9712028}
E.~Witten,
JHEP {\bf 02}, 006 (1998)
hep-th/9712028.

\bibitem{9605150}
A.~Sen,
Nucl. Phys. {\bf B475}, 562 (1996)
hep-th/9605150.

\bibitem{9512059}
A.~Strominger,
Phys. Lett. {\bf B383}, 44 (1996)
hep-th/9512059.

\bibitem{9512062}
P.K.~Townsend,
Phys. Lett. {\bf B373}, 68 (1996)
hep-th/9512062.

\bibitem{AH}
M.F.~Atiyah and N.J.~Hitchin,
Phil. Trans. Roy. Soc. Lond. {\bf A315}, 459 (1985); 
``The Geometry And Dynamics Of Magnetic Monopoles. M.B. Porter Lectures,"
{\it  PRINCETON, USA: UNIV. PR. (1988) 133p}.

\bibitem{9607163}
N.~Seiberg and E.~Witten,
hep-th/9607163.

\bibitem{GIBRUB}
G.W.~Gibbons and P.J.~Ruback,
Commun. Math. Phys. {\bf 115}, 267 (1988).

\bibitem{MANSCH}
N.S.~Manton and B.J.~Schroers,
Ann. Phys. {\bf 225}, 290 (1993).

\bibitem{9402032}
A.~Sen,
Phys. Lett. {\bf B329}, 217 (1994)
hep-th/9402032.

\bibitem{9604070}
A.~Sen,
Nucl. Phys. {\bf B474}, 361 (1996)
hep-th/9604070.

\bibitem{9512196}
K.~Dasgupta and S.~Mukhi,
Nucl. Phys. {\bf B465}, 399 (1996)
hep-th/9512196.

\bibitem{9512219}
E.~Witten,
Nucl. Phys. {\bf B463}, 383 (1996)
hep-th/9512219.


\bibitem{9901159}
P.~Yi,
hep-th/9901159.

\bibitem{9902158}
H.~Awata, S.~Hirano and Y.~Hyakutake,
hep-th/9902158.

\bibitem{9903129}
N.~Kim, S.~Rey and J.~Yee,
JHEP {\bf 04}, 003 (1999)
hep-th/9903129.

\bibitem{9812003}
M.~Gutperle and V.~Periwal,
JHEP {\bf 02}, 018 (1999)
hep-th/9812003.

\bibitem{9807138}
M.~Srednicki,
JHEP {\bf 08}, 005 (1998)
hep-th/9807138.

\bibitem{9804160}
O.~Bergman and B.~Kol,
Nucl. Phys. {\bf B536}, 149 (1998)
hep-th/9804160.

\bibitem{9808073}
M.J.~Strassler, hep-th/9709081, hep-lat/9803009,
hep-th/9808073.

\bibitem{9902181}
I.~Pesando,
``On the effective potential of the Dp - anti-Dp system in type II theories,"
hep-th/9902181.

\bibitem{9903123}
M.~Frau, L.~Gallot, A.~Lerda and P.~Strigazzi,
hep-th/9903123.


\bibitem{POLCAI}
J. Polchinski and Y. Cai, Nucl.Phys. {\bf B296} 91 (1988).

\bibitem{CLNY}
C. Callan, C. Lovelace, C. Nappi and S. Yost, Nucl.Phys. {\bf
B308} 221 (1988).

\bibitem{ONOISH}
T.~Onogi and N.~Ishibashi,
``Conformal Field Theories On Surfaces With Boundaries And Crosscaps,"
Mod. Phys. Lett. {\bf A4}, 161 (1989).

\bibitem{ISHIBASHI}
N.~Ishibashi,
``The Boundary And Crosscap States In Conformal Field Theories,"
Mod. Phys. Lett. {\bf A4}, 251 (1989).

\bibitem{9510227}
J.~Polchinski and A.~Strominger,
``New vacua for type II string theory,"
Phys. Lett. {\bf B388}, 736 (1996)
hep-th/9510227.

\bibitem{9704006}
E.~Gava, K.S.~Narain and M.H.~Sarmadi,
``On the bound states of p-branes and (p+2)-branes,"
Nucl. Phys. {\bf B504}, 214 (1997)
hep-th/9704006.

\bibitem{9708075}
I.~Antoniadis, E.~Gava, K.S.~Narain and T.R.~Taylor,
``Duality in superstring compactifications with magnetic field backgrounds,"
Nucl. Phys. {\bf B511}, 611 (1998)
hep-th/9708075.


\bibitem{BOUNDARY}
O. Bergman and M. Gaberdiel, Nucl.Phys. {\bf B499} 183 (1997) 
hep-th/9701137; \\
M. Li, Nucl.Phys. {\bf B460} 351 (1996) hep-th/9510161; \\
H. Ooguri, Y. Oz and Z. Yin, Nucl.Phys. {\bf B477} 407 (1996) 
hep-th/9606112; \\
K. Becker, M.Becker, D. Morrison, H. Ooguri, Y. Oz and Z. Yin,
Nucl. Phys. {\bf B480} 225 (1996) hep-th/9608116; \\
M. Kato and T. Okada, Nucl. Phys. {\bf B499} 583 (1997) 583
hep-th/9612148; \\
S. Stanciu, hep-th/9708166; \\
A. Recknagel and V. Schomerus, hep-th/9712186; \\
J. Fuchs and C. Schweigert, hep-th/9712257; \\
S. Stanciu and A. Tseytlin, hep-th/9805006; \\
M. Gutperle and Y. Satoh, hep-th/9808080; \\
F. Hussain, R. Iengo, C. Nunez and C. Scrucca, Phys. Lett. {\bf
B409} 101 (1997) hep-th/9706186; \\
M. Bertolini, R. Iengo and C. Scrucca, hep-th/9801110; \\
M. Bertolini, P. Fre, R. Iengo and C. Scrucca, hep-th/9803096; \\
P. Di Vecchia, M. Frau, A. Lerda, I. Pesando, R. Russo and
S. Sciuto, Nucl. Phys. {\bf B507} 259 (1997) hep-th/9707068; \\
M. Billo, P. Di Vecchia, M. Frau, A. Lerda, I. Pesando, R. Russo and
S. Sciuto, Nucl. Phys. {\bf B526} 199 (1998) hep-th/9802088; 
hep-th/9805091.

\bibitem{WITTBFT}
E.~Witten, Nucl. Phys. {\bf B268} 253 (1986).


\bibitem{WITTSFT}
E.~Witten, Nucl. Phys. {\bf B276} 291 (1986).


\bibitem{CUBIC}
A.~Strominger, G.~T.~Horowitz, J.~Lykken and R.~Rohm, Phys. Rev. Lett.
{\bf 57} 283 (1986).

\bibitem{9402113}
C.G.~Callan, I.R.~Klebanov, A.W.~Ludwig and J.M.~Maldacena,
Nucl. Phys. {\bf B422}, 417 (1994)
hep-th/9402113.

\bibitem{9404008}
J.~Polchinski and L.~Thorlacius,
Phys. Rev. {\bf D50}, 622 (1994)
hep-th/9404008.

\bibitem{9811237}
A.~Recknagel and V.~Schomerus,
hep-th/9811237.

\bibitem{9807161}
S.~Elitzur, E.~Rabinovici and G.~Sarkisian,
``On least action D-branes,"
Nucl. Phys. {\bf B541}, 246 (1999)
hep-th/9807161.

\bibitem{9902105}
A.~Sen,
hep-th/9902105.

\bibitem{FMS}
D.~Friedan, E.~Martinec and S.~Shenker,
Nucl. Phys. {\bf B271}, 93 (1986).


\end{thebibliography}
\end{document}